\iftrue
\documentclass[aps,prl,twocolumn,
			   groupedaddress,superscriptaddress,
			   amsfonts,amssymb,amsmath,
			   citeautoscript,
			   a4paper]{revtex4-1}
\else
\documentclass[aps,pra,preprint,
			   groupedaddress,superscriptaddress,notitlepage,
			   amsfonts,amssymb,amsmath,
			   citeautoscript,
			   a4paper]{revtex4-1}
\fi

\usepackage[pdftex]{hyperref}
\hypersetup{colorlinks,
			linkcolor={blue!75!black!80!yellow},
			citecolor={blue!75!black!80!yellow},
			urlcolor={blue!75!black!80!yellow},
			pdfstartview=FitH}
\usepackage{graphicx}
\usepackage{mathrsfs}
\usepackage{amsthm}
\usepackage{physics}
\usepackage{xspace}
\usepackage{braket}
\usepackage{xr}
\usepackage{gensymb} % to use \degree command
\usepackage{xcolor,soul}
\usepackage{stmaryrd} % provides \llbracket and \rrbracket for difference-operations
\usepackage[UKenglish]{babel}
\usepackage{placeins} % provides \FloatBarrier
\usepackage{siunitx}
\DeclareSIUnit[number-unit-product=]\percent{\char`\%} % remove space before percentage "units"
\usepackage{makecell}
\usepackage{diagbox}
\usepackage{float}
\usepackage{amsmath} % for 'bmatrix' environment
\usepackage{empheq} % for numbering different cases in an equation
\usepackage{physics}
\newcommand{\iu}{\mathrm{i}}
\newcommand{\expn}{\mathrm{e}}

\newcommand{\rmred}{\mathrm{r}}
\newcommand{\rmgreen}{\mathrm{g}}
\newcommand{\rmblue}{\mathrm{b}}
\usepackage{txfonts}  %Times Roman fonts
\usepackage{txfontsb} %Addition for txfonts, including old style numerals and greek

\newcommand*\diff{\mathop{}\mathrm{d}}

\newcommand{\appropto}{\mathrel{\vcenter{
			\offinterlineskip\halign{\hfil$##$\cr
				\propto\cr\noalign{\kern2pt}\sim\cr\noalign{\kern-2pt}}}}}

\makeatletter
\newcommand*{\addFileDependency}[1]{% argument=file name and extension
  \typeout{(#1)}% latexmk will find this if $recorder=0 (however, in that case, it will ignore #1 if it is a .aux or .pdf file etc and it exists! if it doesn't exist, it will appear in the list of dependents regardless)
  \@addtofilelist{#1}% if you want it to appear in \listfiles, not really necessary and latexmk doesn't use this
  \IfFileExists{#1}{}{\typeout{No file #1.}}% latexmk will find this message if #1 doesn't exist (yet)
}
\makeatother

\newcommand*{\myexternaldocument}[1]{%
    \externaldocument{#1}%
    \addFileDependency{#1.tex}%
    \addFileDependency{#1.aux}%
}

\myexternaldocument{SM_draft1}

\usepackage{scalerel}

\usepackage{textcomp} % for \textrightarrow
\usepackage{xifthen}
\usepackage{etoolbox}
\newboolean{togglecomments}
\newboolean{togglechanges} 

\setboolean{togglecomments}{true}
\setboolean{togglechanges}{false}

\newcommand{\comment}[2]{%
    \ifbool{togglecomments}%
    {\textcolor{blue!70!black}{\small\textsf{%
    \textsuperscript{\textsc{\textsf{\MakeLowercase{#1}}}}%
    [#2]}}} % if true, show comments
    {}}     % if false, do nothing
\newcommand{\swap}[2]{\ifbool{togglechanges}
    {#2}  % updates-only version
    {\textcolor{red!70!black}{[#1]}\textrightarrow{}\textcolor{green!50!black}{[#2]}}}
\newcommand{\remove}[1]{\ifbool{togglechanges}
    {}    % updates-only version
    {\textcolor{red!70!black}{#1}}}
\newcommand{\inset}[1]{\ifbool{togglechanges}
    {#1}  % updates-only version
    {\textcolor{green!50!black}{#1}}}
\newcommand{\optional}[1]{\ifbool{togglechanges}
    {#1}  % updates-only version
    {\textcolor{yellow!50!orange!80!gray}{#1}}}
\newcommand{\citeremind}[1]{%
    [\textcolor{blue!75!black!80!yellow}{
        $\blacksquare$%
           \ifthenelse{\isempty{#1}}
               {}
               {\textsuperscript{\textsf{#1}}}%
        }]\xspace}
\newcommand{\todo}[1]{% don't use in captions/subfloats etc.
    \textcolor{orange!80!yellow!95!black}{\textbf{[}%
        \ifthenelse{\isempty{#1}}%
        {\text{$\blacksquare$}}%
        {{\small\textsf{#1}}}%
        \textbf{]}}}
        
\newcommand{\hkuaffil}{\footnotesize Department of Physics and HK Institute of Quantum Science and Technology, The University of Hong Kong, Pokfulam, Hong Kong, China}

\begin{document}

\title{Synthetic SU(3) non-Abelian gauge fields in photonic synthetic frequency dimensions}

\author{Bengy Tsz Tsun Wong}
\affiliation{\hkuaffil}
\author{Shu Yang}
\affiliation{\hkuaffil}
\author{Zehai Pang}
\affiliation{\hkuaffil}
\author{Yi~Yang}
\email{yiyg@hku.hk}
\affiliation{\hkuaffil}

\begin{abstract}

Synthetic gauge fields provide a powerful route to engineer photonic wave dynamics, especially in synthetic frequency dimensions formed by modulated resonator modes. While most previous realizations have focused on Abelian or lower-rank non-Abelian structures, a photonic implementation of SU(3) gauge fields in the synthetic frequency dimension has remained lacking. Here we propose a route to realize SU(3) gauge structure using three coupled ring resonators with modulation- and interference-engineered mode conversion. We formulate a tight-binding Hamiltonian with matrix-valued hopping generated by the Gell-Mann matrices and show how its evolution can be decomposed into experimentally implementable optical elements. The resulting platform supports SU(3)-structured band dispersions with color-resolved texture and wave-packet dynamics shaped by SU(3) gauge fields. Full-wave-inspired numerical modeling further confirms the expected spectral response and Zitterbewegung-like motion in SU(3) subspaces. Our results introduce SU(3) gauge fields into synthetic frequency photonics and establish a route toward higher-rank non-Abelian control of light in modulated resonator systems.

\end{abstract}

\maketitle

Synthetic gauge fields have become a powerful concept in photonics for engineering wave transport beyond the constraints of natural material response~\cite{aidelsburger2018review,ding2025progress,song2025artificial}. By encoding effective phases and couplings into optical evolution, synthetic gauge fields enable controllable band deformation, directional transport, topological effects, and gauge-dependent wave dynamics in systems that are otherwise electrically neutral~\cite{zhai2015degenerate,lin2011spin,zhang2015spin,goldman2014light,yang20202d,sedov2018zitterbewegung,hasan2022wave,polimeno2021experimental,lovett2023observation,chen2019non,cheng2023artificial,pang2024synthetic,wong2025synthetic}. In this way, photonic platforms provide an especially flexible setting in which gauge structures can be designed and reconfigured through modulation, coupling architecture, and synthetic coordinates.

A particularly fruitful development in this direction is the use of synthetic dimensions~\cite{celi2014synthetic,yuan2018synthetic,ozawa2019topological,lustig2021topological,yuan2021synthetic,arguello2024synthetic}, where a wide range of modal degrees of freedom are promoted to lattice coordinates for light. Among the various implementations, the synthetic frequency dimension has emerged as a versatile platform because dynamic modulation can couple equally spaced resonant modes of ring resonators and thereby generate tight-binding lattices directly in frequency space~\cite{hey2018advances,ozawa2016synthetic,yuan2016bloch,bell2017spectral,qin2018spectrum,dutt2020single,balvcytis2022synthetic,tusnin2020nonlinear,suh2024photonic,du2022giant,wang2021generating,cheng2021arbitrary,hu2020realization,senanian2023programmable}. This approach has enabled a broad range of photonic functionalities, including controlled spectral transport, band engineering, and higher-dimensional synthetic lattices, all within reconfigurable or even integrated optical architectures~\cite{dutt2019experimental,dutt2022creating,senanian2023programmable,cheng2023multi,balvcytis2022synthetic}.

Within synthetic frequency dimensions, synthetic gauge fields arise naturally through modulation phases and intermode coupling phases. Most previous studies have focused on Abelian gauge structures \cite{yuan2016photonic,ozawa2017synthetic,yuan2019photonic,dutt2019experimental,lustig2019photonic,balvcytis2022synthetic,dinh2024reconfigurable}, where the accumulated phase is described by a commuting U(1) field. More recently, attention has turned to non-Abelian synthetic gauge fields~\cite{chen2019non,cheng2023artificial,neef2023three,yang2024non,wong2025synthetic,cheng2025non,liu2025general,dong2025sl, yang2026non}, which introduce internal degrees of freedom and matrix-valued couplings whose order matters. In photonics, such non-Abelian structures provide access to richer mode conversion, spin-orbit-like coupling, matrix gauge transport, and multicomponent band textures that have no counterpart in purely scalar synthetic lattices.

Despite this progress, photonic realizations of non-Abelian gauge fields in the synthetic frequency dimension have so far remained largely restricted to lower-rank structures, while the full SU(3) gauge group and higher-rank structures are still lacking. Compared with SU(2), the SU(3) gauge group contains eight generators and supports a substantially enlarged internal structure, allowing three-component coupling, nontrivial color texture, and more complex gauge-dependent dynamics. Establishing an SU(3) photonic platform would therefore extend the accessible non-Abelian physics in synthetic dimensions and provide a route to qualitatively new forms of spectral and dynamical control.

In this work, we propose a scheme to realize SU(3) gauge fields in a photonic synthetic frequency dimension using three coupled ring resonators with modulated and interferometrically engineered mode conversion. Starting from a tight-binding Hamiltonian with matrix-valued hopping governed by the Gell-Mann generators, we show how the corresponding evolution can be decomposed into experimentally implementable optical elements, including phase retarders, rotators, and couplers. This construction yields a photonic route to the full SU(3) structure and, in reduced form, to experimentally accessible partial SU(3) subspaces. We further analyze the resulting band structures and associated colour textures, identify transmission signatures in the three projected channels, and examine wave-packet dynamics induced by the non-Abelian gauge fields. Our results establish synthetic frequency photonics as a platform for accessing SU(3) gauge structure and broaden the scope of non-Abelian control in time-modulated optical systems.

\begin{figure}
    \centering
    \includegraphics[width=\linewidth]{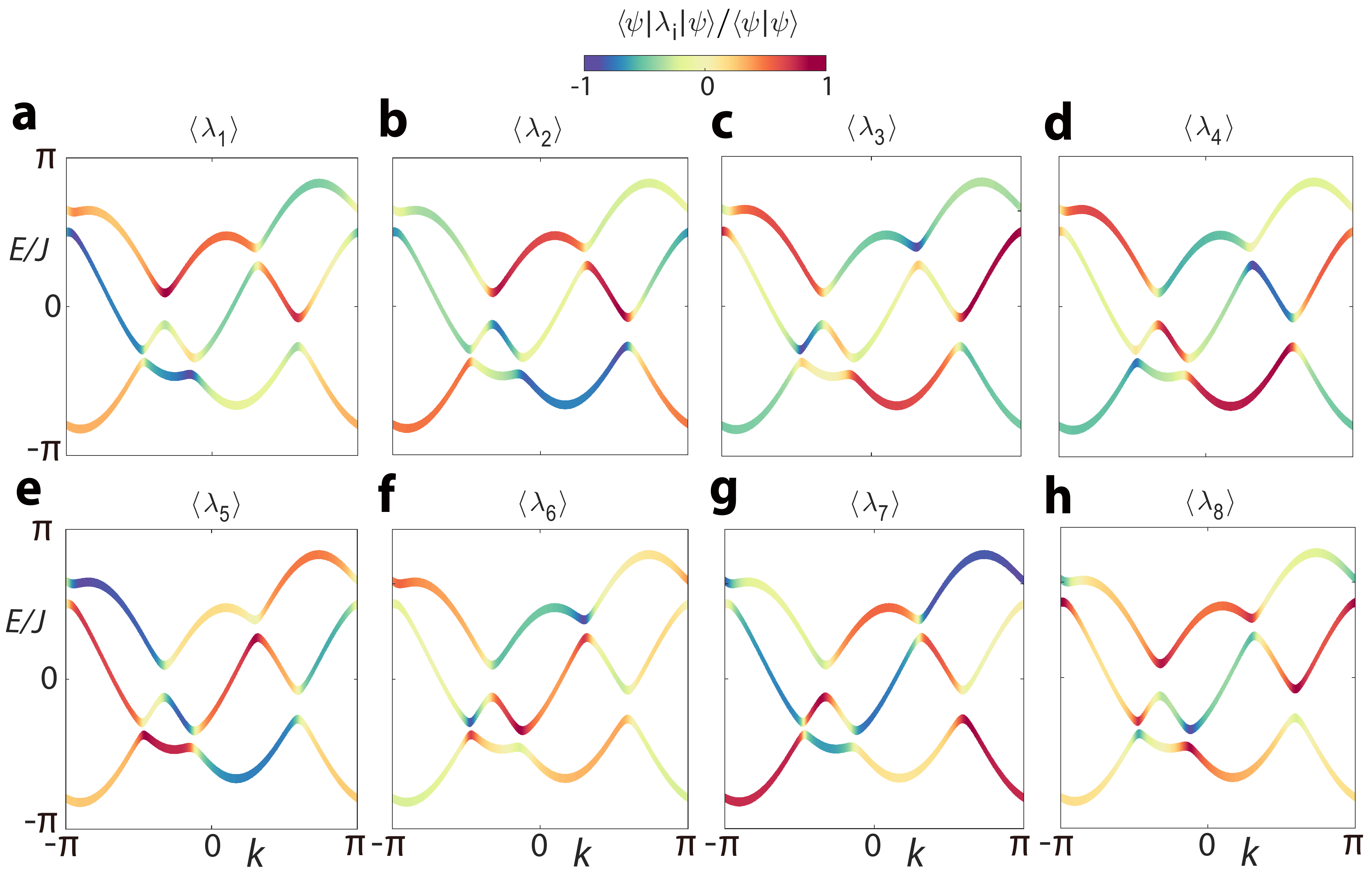}
    \centering
    \caption{\textbf{Photonic band structure of synthetic $\mathrm{SU(3)}$ gauge group and colour texture.} 
    \textbf{a.-h.} Dispersion relations of three energy bands with opening band gaps. They showcase the colour texture of the full $\mathrm{SU(3)}$ group given by $\langle \lambda_i\rangle = \langle \psi | \lambda_i|\psi\rangle /\langle \psi | \psi\rangle$, where $i=1,2,\cdots, 8$. The eigenstates are $|\psi\rangle=\{\psi_1, \psi_2 , \psi_3 \}$, where we use $\psi_1$ to label the lowest band, $\psi_2$ for the middle band, and $\psi_3$ for the highest band, respectively. 
    The gauge-field parameters chosen are we take the numerical values of gauge fields as $\{\theta_1, \theta_2, \theta_3, \theta_4,\theta_5, \theta_6,\theta_7, \theta_8\}=  \{0.6, -0.3, 1.2 , 0.8,-1.6 , 0.2, 0.9, 0.4    \} \,\mathrm{rad}$. We also set $\omega_7/J = 0.7$ for all figures. 
    } 
    \label{fig:SU3Band_colourtexture}
\end{figure}

\emph{Hamiltonian formulation.}
To realize the tight-binding Hamiltonian featuring SU(3) gauge fields in the photonic synthetic frequency dimension, we consider a general Hamiltonian 
\begin{align} \label{eq:HamFullSU3}
\hat{H}_0 = J \sum_n \hat{a}^\dagger_{n+1}  \expn^{\iu (\theta_1 \lambda_1 + \theta_2 \lambda_2 + \cdots +\theta_8\lambda_8)} \hat{a}_{n} + \mathrm{h.c.}\,,
\end{align} 
where $\{\lambda_{i}\}_{i=1}^8$ are the eight Gell-Mann matrices, i.e. generators satisfying the $\mathrm{SU(3)}$ Lie algebra of $[\lambda^a , \lambda^b] = 2\iu f^{abc}\lambda^c$ with $f^{abc}$ the totally antisymmetric structural constant, and $a_n = (a_{n\mathrm{r}}, a_{n\mathrm{g}}, a_{n\mathrm{b}})^T$ where {r,g,b} stands for red, green, blue as the internal colour degree of freedom as used in the nomenclature of quantum chromodynamics. 
%
% The Bloch Hamiltonian in the reciprocal space is
% \begin{align} \label{eq:HamSU3momentum}
% H = 2J\sum_k  \hat{a}^\dagger_{k} \cos(k +\theta_1\lambda_1 +\cdots+ \theta_8\lambda
% _8)  \hat{a}_{k}  \,,
% \end{align}
% which can be explicitly expressed as
% \begin{equation}
% \begin{aligned}
% H &=2J  \sum_{j=1}^3\sum_k  \hat{a}_{k}^\dagger\bigg[P_j\cos(k +z_j |\theta|)I +\cos(k + z_j |\theta|) \\
% &\quad\quad\quad\quad\quad\quad\quad\times
% \bigg(Q_j \frac{\theta_c}{|\theta|} +R_j d^{abc}\frac{\theta_a \theta_b}{|\theta|^2}  \bigg)\lambda_c \bigg]_{\alpha\beta} \hat{a}_{k} \,,
% \end{aligned}
% \end{equation}
By utilizing the linearization scheme of the SU(3) group element~\cite{kaiser2022solving}, we express the Bloch Hamiltonian in reciprocal space as
\begin{equation}
\begin{aligned}
\hat{H}_0(k) &=2J  \sum_{j=1}^3\sum_k  \hat{a}_{k}^\dagger\bigg[P_j\cos(k +z_j \|\theta\|)I +\cos(k + z_j \|\theta\|) \\
&\quad\quad\quad\quad\quad\quad\quad\times
\bigg(Q_j \frac{\theta_c}{\|\theta\|} +R_j d^{abc}\frac{\theta_a \theta_b}{\|\theta\|^2}  \bigg)\lambda_c \bigg] \hat{a}_{k} \,,
\end{aligned}
\end{equation}
where $\|\theta\| = \sqrt{\theta_1^2 + \theta_2^2 +\cdots + \theta_8^2}$, $d_{abc}$ is the totally symmetric structural constant of $\mathrm{SU}(3)$ and Einstein's summation convention is adopted on the repeated indices $a,b,c$; $P_j = (z_j^2 -1/3)/(3z_j^2 -1)=1/3\,,\quad Q_j= z_j/(3z_j^2-1)\,\quad \text{and}\,\quad R_j = 1/(3z_j^2 -1)$ with $z_1, z_2, z_3$ satisfying the constraint of $z_1 + z_2 + z_3 =0$ (details of derivation see SM Sec.\,1). 
To enable the color-orbit coupling (COC)~\cite{barnett20123,madasu2025experimental}[similar to the spin-orbit coupling in SU(2)], we can further introduce an additional on-site potential in the Hamiltonian via $\hat{H}=\hat{H}_0+\hat{H}_{\mathrm{coc}}$, where $\hat{H}_{\mathrm{coc}}= -\sum_{n}\omega_7 \hat{a}^\dagger_n \lambda_7 \hat{a}_n$ and $\omega_7$ is the strength of the on-site potential.
An example band structure overlaid by the corresponding color texture projected onto the $\{\lambda_{i}\}_{i=1}^8$ basis is plotted in Fig.\ref{fig:SU3Band_colourtexture}, where color texture exchanges happen at the avoided crossings among the three bands.

% \comment{yy}{to be revised}
% \comment{yy}{why not show the full lambda1-8 band here?}\comment{BW}{I thought you said we don't need to change figure 1 yesterday, I put the full group colour texture in SM. Here we only have lambda1,2 and lambda8 here}\comment{yy}{let us show the full stuff in the main and move the partial stuff in the SM after the derivation of the partial scheme}\comment{BW}{OKay, done.}

\begin{figure}
    \centering
    \includegraphics[width=\linewidth]{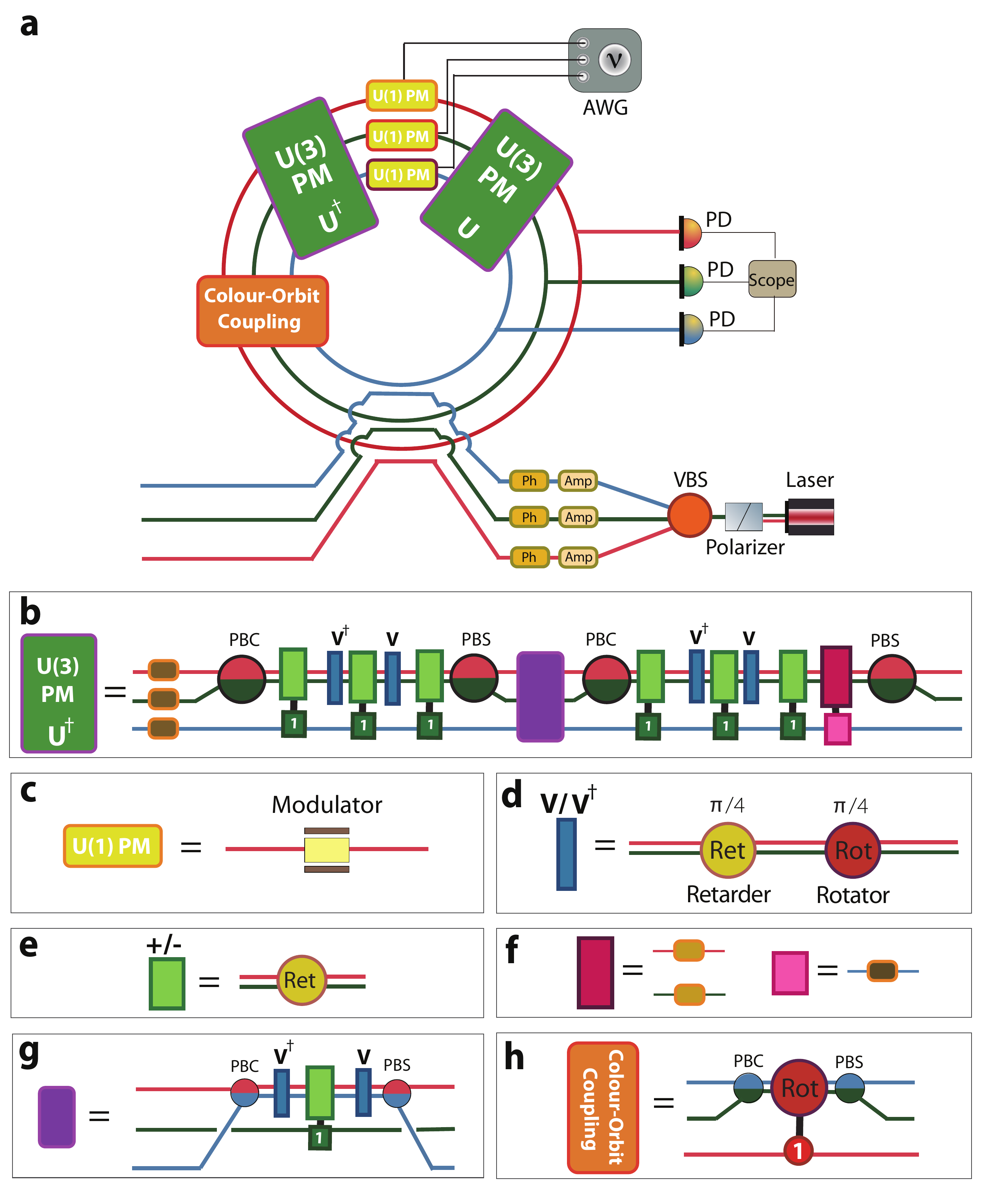}
    \centering
    \caption{\textbf{Setup schematic for realizing the full $\mathrm{SU(3)}$ gauge group in the synthetic frequency dimension of ring resonators.} 
    \textbf{a.} Generic setup. 
    % \comment{yy}{PBS/C is infact a 2x1 port device. Now it looks 2x2.} \comment{BW}{I update both figures above and SM}
    \textbf{b.} The $\mathrm{U(3)}$ matrix decomposition equipped with PBS/C. PBS/C: Polarization beam splitter/combiner. From left to right, the three small brown boxes describe three identical $\mathrm{U(1)}$ phase change $\expn^{\iu\delta}$.
    The green rectangle composite represents  $ \expn^{\iu\alpha\lambda_3}=\expn^{\iu\alpha \sigma_z}\oplus1$, of which the retarder introduces phase delay in the first two modes (red ring and green ring) with the third mode (blue ring) being unaffected (see Fig. $\mathbf{e}$). 
    The green rectangle sandwiched with two bluish-green rectangle represents $\expn^{\iu\beta\lambda_2}= V \expn^{\iu\beta\sigma_z}V^\dagger \oplus 1$ is a $\mathrm{SU(2)}$ phase change (see panel~\textbf{d} and \textbf{e}). The fourth green rectangle composite is $\expn^{\iu\gamma\lambda_3}$, which is a phase retardation similar to the second element but with a different phase parameter $\gamma$. The fifth element $\expn^{\iu \theta\lambda_5}=V^\prime (\expn^{\iu\theta}\oplus 1 \oplus \expn^{-\iu\theta}) V^{\prime\dagger}$ illustrated by the purple rectangular box couples the first and the third mode (red ring and blue ring) respectively, which involves a controlled mixture in the configuration enabled by polarization beam combiner (PBC) and polarization beam splitter (PBS) (see panel~\textbf{g}).
    The sixth element is again a retarder composite with a different parameter $a$. The seventh element is $\expn^{\iu b\lambda_2}$, which is similar to the third element but with a different phase parameter $b$. The eighth element $\expn^{\iu c\lambda_3}$ is again a retarder composite but with a different retarded phase parameter $c$. 
    Finally, the last element $\expn^{\iu\phi\lambda_8}=\expn^{\iu (\phi/\sqrt{3}) I} \oplus \expn^{-\iu 2\phi/\sqrt{3}}$, a red-rectangle connected to a pink square represents two separated $\mathrm{U(1)}$ phase change with different magnitude (see panel~\textbf{f}).  
    \textbf{d.-g.} The sub-elements in the $\mathrm{U(3)}$ matrix composite. In between the $U^\dagger$ and $U$ matrices lies the diagonalized transfer matrix $\tilde{\mathbf{T}} =\mathrm{diag}( \expn^{-\iu 2Jt_\mathrm{R}\cos(k+\mu_1)},  \expn^{-\iu 2Jt_\mathrm{R}\cos(k+\mu_2)},  \expn^{-\iu 2Jt_\mathrm{R}\cos(k+\mu_3) })$, which generates three different, separate $\mathrm{U(1)}$ modulation for the three different modes (see panel~\textbf{c}). For all cases, the item with $``\oplus 1"$ means the laser light passes through the ring without alteration. \textbf{h.} The colour-oribit coupler $\mathbf{T}_{\mathrm{coup}}=\expn^{\iu\tilde{\phi}\lambda_7}=1\oplus \expn^{\iu\tilde{\phi}\sigma_y}$. PD: Photodetector. VBS: Variable beam splitter. Amp and Ph: Amplitude modulator and phase retarder. AWG: Alternative-wave generator.}  
    \label{fig:setupschematic}
\end{figure}

\emph{From transfer matrix to optical realization.} 
To realize the Block Hamiltonian in its full form $H(k) = 2J\cos(k+\theta_1 \lambda_1 + \theta_2 \lambda_2 + \cdots + \theta_8 \lambda_8) - \omega_7 \lambda_7$ (see more analytical discussions in SM. Sec.\,2), we first consider diagonalizing the first part of the Hamiltonian; this yields
\begin{equation}
\begin{aligned}
\tilde{H}_0(k) &= U^\dagger H_0(k) U \\
&= \begin{pmatrix}
 2J\cos(k+\mu_1) & 0 & 0\\
 0 &  2J\cos(k+\mu_2) & 0 \\
 0   &   0  & 2J\cos(k+\mu_3)\\
\end{pmatrix} \,,
\end{aligned}
\end{equation}
where $
\mu_j = (2/\sqrt{3}) \| \theta \| \cos(\phi - 2\pi(j-1)/3 )
$ for $j=1,2,3$ with $ \phi = (1/3)\cos^{-1}(3\sqrt{3}\det X/2 \|\theta \|^3 )$  and $ \det X = (2/3)\sum_{a,b,c =1}^8 d_{abc }\theta_a \theta_b \theta_c$ with $X=\theta_1 \lambda_1 + \theta_2 \lambda_2 + \cdots + \theta_8 \lambda_8$ the Lie vector, and $d_{abc}$ is the totally symmetric constant in $\mathrm{SU(3)}$. Since our $\mathrm{SU(3)}$ Hamiltonian is Hermitian, it can be diagonalized by a unitary matrix of dimension three, i.e., a $\mathrm{U(3)}$ matrix. In addition, as any $\mathrm{U(3)}$ element can be decomposed into nine exponential factors with their nine respective Euler angles, given by Ref.~\cite{byrd1998differential},
\begin{equation}
D(\delta, \alpha,\beta,\gamma,\theta,a,b,c,\phi) = \expn^{\iu \delta}\expn^{\iu \alpha \lambda_3} \expn^{\iu \beta \lambda_2}\expn^{\iu \gamma \lambda_3}\expn^{\iu \theta \lambda_5} \expn^{\iu a\lambda_3} \expn^{\iu b\lambda_2}  \expn^{\iu c\lambda_3} \expn^{\iu \phi\lambda_8}\,.
\end{equation}
We can map $U = D(\delta, \alpha,\beta,\gamma,\theta,a,b,c,\phi)$. 
To derive the transfer matrix for optical realization, consider the evolution of an initial state $\psi_n(t)$ such that $\psi_n^\prime (t+t_\mathrm{R}) = \expn^{-\iu H_0 t_\mathrm{R}}\psi_n (t)= \mathbf{T}\psi_n (t)$, where the transfer matrix is given by $\mathbf{T}= \expn^{-\iu H_0 (k) t_\mathrm{R}}= U \expn^{-\iu\tilde{H}_0(k) t_\mathrm{R}} U^\dagger = D \mathbf{\tilde{T}}  D^\dagger$ and $t_{\mathrm{R}}$ is the round-trip time. The diagonalized transfer matrix is $\tilde{\mathbf{T}} =\mathrm{diag}( \expn^{-\iu 2Jt_\mathrm{R}\cos(k+\mu_1)},  \expn^{-\iu 2Jt_\mathrm{R}\cos(k+\mu_2)},  \expn^{-\iu 2Jt_\mathrm{R}\cos(k+\mu_3)} ) $. Since each exponential factor in $D$ can be either further decomposed as a direct sum (for $\lambda_3 $ and $\lambda_8$) or further be diagonalized (for $\lambda_2$ and $\lambda_5$), the $\mathrm{U(3)}$ element can be fully decomposed as follows:
\begin{equation}
\begin{aligned}
U &= \expn^{\iu\delta} (\expn^{\iu \alpha\sigma_z}\oplus 1 ) (V \expn^{\iu\beta \sigma_z} V^\dagger \oplus 1)(\expn^{\iu \gamma \sigma_z}\oplus 1 ) [V^\prime (\expn^{\iu\theta}\oplus 1 \oplus \expn^{-\iu\theta}) V^{\prime\dagger}]\\
&\quad\quad\quad\quad\times  (\expn^{\iu a\sigma_z}\oplus 1 ) (V \expn^{\iu b\sigma_z} V^\dagger \oplus 1 ) (\expn^{\iu c\sigma_z}\oplus 1 ) \left(\expn^{\iu\frac{\phi}{\sqrt{3}} I} \oplus \expn^{-\iu\frac{2\phi}{\sqrt{3}}}\right)\,,
\end{aligned}
\end{equation}
where $V= R_z \big( \frac{\pi}{2} \big) R_y \big(\frac{\pi}{2}\big) = \expn^{-\iu \frac{\pi}{4}\sigma_z } \expn^{ -\iu \frac{\pi}{4}\sigma_y }$ and
\begin{equation}
V^\prime = \begin{pmatrix}
\expn^{-\iu\frac{\pi}{4}} & 0 & 0\\
0 & 1 & 0\\
0 & 0 & \expn^{\iu\frac{\pi}{4}}\\
\end{pmatrix}
\begin{pmatrix}
\cos \frac{\pi}{4} & 0 & \sin\frac{\pi}{4} \\
0 & 1 & 0 \\
- \sin\frac{\pi}{4} & 0 & \cos\frac{\pi}{4} \\
\end{pmatrix}\,,
\end{equation}
which is a rotation followed by a retardation acting on the first mode (red ring in Fig.~\,\ref{fig:setupschematic}) and the third mode (blue ring). Hence, the transfer matrix $\mathbf{T}$ is solely composed of different combinations of optical rotators and retarders in different orders. To enable colour-orbit coupling, we further implement the coupler $\mathbf{T}_{\mathrm{coc}}=\expn^{\iu \tilde{\phi}\lambda_7}$. It introduces a rotational coupling between the blue and the green rings, so that the full transfer matrix is given by $\mathbf{T}_{\mathrm{tot}}=\mathbf{T}\cdot \mathbf{T}_{\mathrm{coc}}$. The photonic realization is illustrated in Fig.~\ref{fig:setupschematic}. The details of the derivation are in SM Sec.\,3.

Three single-mode laser beams, from the same laser,  enter the three ring resonators  (labeled by red, green, and blue, respectively) and propagate clockwise in the rings.
Before entering the three rings, arbitrary SU(3) input states can be prepared via a variable beam splitter (VBS), amplitude amplifiers, and phase retarders.
We use three polarization-maintaining fiber rings to encode the SU(3) internal degrees of freedom.
In the absence of inter-ring coupling, each ring occupies only a single polarization, say, the slow axis.
The mutual coupling among the rings is realized via polarization beam splitters/combiners (PBS/Cs).  A PBC is used such that optical waves from two rings occupy two orthogonal polarizations; after controlled couplings are implemented, we use a PBS to split the orthogonal polarizations into the slow axis of two separate rings.

As the beams enter the rings, they first pass through the coupler $\mathbf{T}_{\mathrm{coc}}=\expn^{\iu \tilde{\phi}\lambda_7}= 1\oplus \expn^{\iu\tilde{\phi}\sigma_y}$, which couples to the second and the third mode (green mode and blue mode) respectively; this coupler provides the color-orbit coupling.
It follows to enter the $\mathrm{U(3)}$ matrix composite following the order of $ \expn^{\iu\delta} \expn^{\iu \alpha \lambda_3} \expn^{\iu \beta \lambda_2}\expn^{\iu \gamma \lambda_3} \expn^{i \theta \lambda_5} \expn^{\iu a\lambda_3} \expn^{\iu b\lambda_2}  \expn^{\iu c\lambda_3} \expn^{\iu\phi\lambda_8}$ one by one. In particular, $ \expn^{\iu\alpha\lambda_3}=\expn^{\iu\alpha \sigma_z}\oplus1$ introduces the retarded phase in the first ring (red ring) and the second ring (green ring); next, $\expn^{i\beta\lambda_2}= V \expn^{\iu\beta\sigma_z}V^\dagger \oplus 1 =\expn^{-\iu \frac{\pi}{4}\sigma_z } \expn^{ -\iu \frac{\pi}{4}\sigma_y}  \expn^{\iu\beta\sigma_z}     \expn^{\iu \frac{\pi}{4}\sigma_z } \expn^{\iu \frac{\pi}{4}\sigma_y}   \oplus 1 $ which couples the first ring (red ring) and second ring (green ring) with sandwiched $\pi/4$ phase rotation with retardation in between. The fifth element, $\expn^{\iu\theta\lambda_5} =V^\prime (\expn^{\iu\theta}\oplus 1 \oplus \expn^{-\iu\theta}) V^{\prime\dagger}$ is a coupler which has a similar functionality as $ \expn^{i \beta \lambda_2}$, except that it couples the red ring and the blue ring instead of the red ring and the green ring. 
In the optical design, it amounts to giving a controlled mixture with the aid of a PBC (polarization beam combiner) and a PBS (polarization beam splitter). After passing through the color-orbit coupler and $U^\dagger$ matrix composite, the three single-mode laser lights pass through three different $\mathrm{U(1)}$ modulations, with the red ring experiencing 
$\expn^{-\iu 2Jt_\mathrm{R}\cos(k+\mu_1)}$, the blue ring experiencing $\expn^{-\iu 2Jt_\mathrm{R}\cos(k+\mu_2)}$ and the green ring passing through $\expn^{-\iu 2Jt_\mathrm{R}\cos(k+\mu_3)}$ respectively. Afterwards, they pass through the $\mathrm{U(3)}$ matrix composite and are detected by the photodetectors (PDs).

\begin{figure}
    \centering
    \includegraphics[width=\linewidth]{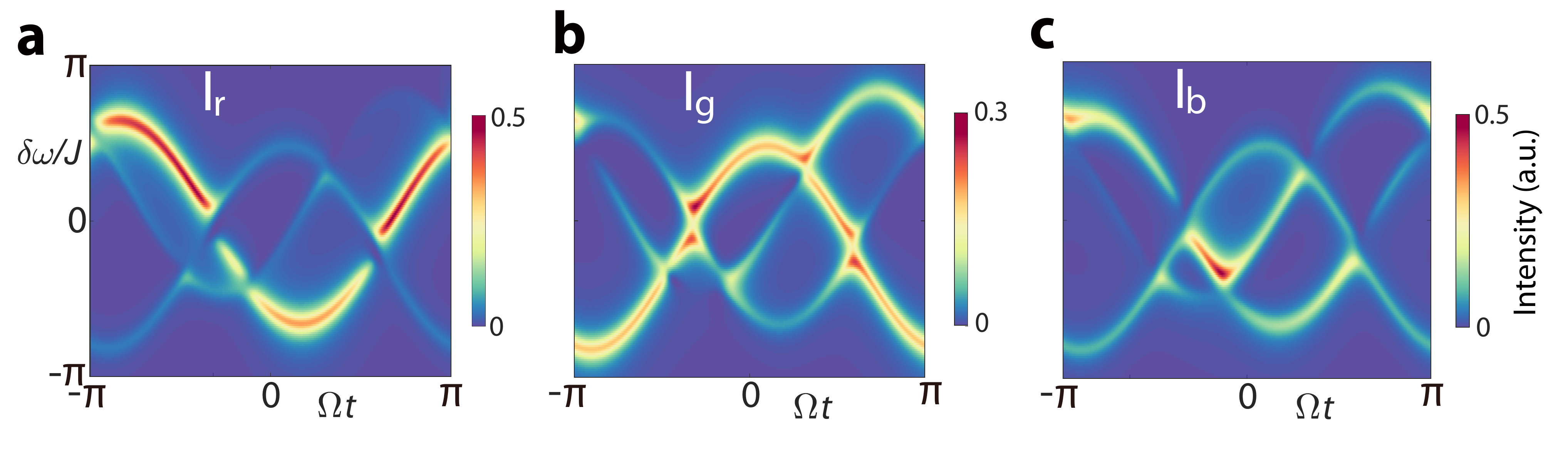}
    \centering
    \caption{
    \textbf{Projected intensities $I_\rmred$ (a), $I_\rmgreen$ (b) and $I_\rmblue$ (c) from each each ring.}
    The gauge-field parameters are chosen to be the same as in Fig.\ref{fig:SU3Band_colourtexture} and in addition we also pick $\gamma_{\mathrm{in}}/J = 0.2$. The quasi-momentum $k$ is equal to $\Omega t$ in an actual experiment. 
    }
    \label{fig:SU3signal}
\end{figure}

\emph{Band dispersion and transmission spectroscopy.} In an experiment, the transmission intensities of the band structure with three degrees of freedom $I_{\mathrm{r}}(\delta\omega,k)$, $I_{\mathrm{g}}(\delta\omega,k)$ and $I_{\mathrm{b}}(\delta\omega,k)$ can be probed via transmission spectroscopy along the synthetic dimension. The expected signal detected by the photo-detector (PD) via drop ports of the rings can be obtained by the following equation: 
\begin{align} \label{eq:projectedIntensity}
I_\alpha  = |\psi_\alpha|^2 =\bigg\vert   \mathrm{i}\gamma_{\mathrm{in}} \sum_{j=1}^3 \frac{ \braket{\psi_\alpha | \psi_j } \braket{ \psi_j |\psi_{\mathrm{\mathrm{i}n}}}} { E_j -\delta\omega - \iu\gamma_{\mathrm{\mathrm{i}n}} } \bigg\vert^2 \,,
\end{align}
where $\alpha=\left\{\mathrm{r},\mathrm{g},\mathrm{b}\right\}$ are the three projected colored modes representing the red, green and blue states, $\ket{\psi_{1,2,3}}$ are the eigenstates of the full Bloch Hamiltonian together with the on-site potential, $\delta\omega$ is the laser detuning, $\ket{\psi_{\mathrm{\mathrm{i}n}}}$ is the input state and $\gamma_{\mathrm{in}}$ is the weak coupling to the waveguide. The transmission intensity in each of the rings is detected by each photodetector, as plotted in Fig.~\ref{fig:SU3signal}; these intensities follow the band structure outlined in Fig.~\ref{fig:SU3Band_colourtexture}.

Due to the complexity of the $\mathrm{SU(3)}$ group, an analytical closed-form mapping between the eight gauge-field parameters $\{\theta_1, \theta_2, \theta_3, \theta_4,\theta_5, \theta_6,\theta_7, \theta_8\}$ and the nine Euler-angle parameters $\{\delta, \alpha,\beta,\gamma,\theta, a,b,c,\phi\}$ is formidable. 
Nevertheless, it is possible to obtain an explicit analytical mapping if we consider only a subset of the SU(3) group so that the transfer matrix $\mathbf{T}= \expn^{-\iu H (k) t_\mathrm{R}}$ can be directly mapped onto the $\mathrm{SU(3)}$ matrix elements. This can be achieved, for example, by considering only three generators $\{\lambda_1,\lambda_2,\lambda_8 \}$ and three corresponding gauge-field parameters $\{\theta_1,\theta_2,\theta_8 \}$. 
The details of this derivation and simplification, with a complete setup schematic and numerical verifications, are described in SM Sec. 4-6. 
The dynamics analysis below also focuses on this simplified scheme.

\begin{figure}
    \centering
    \includegraphics[width=\linewidth]{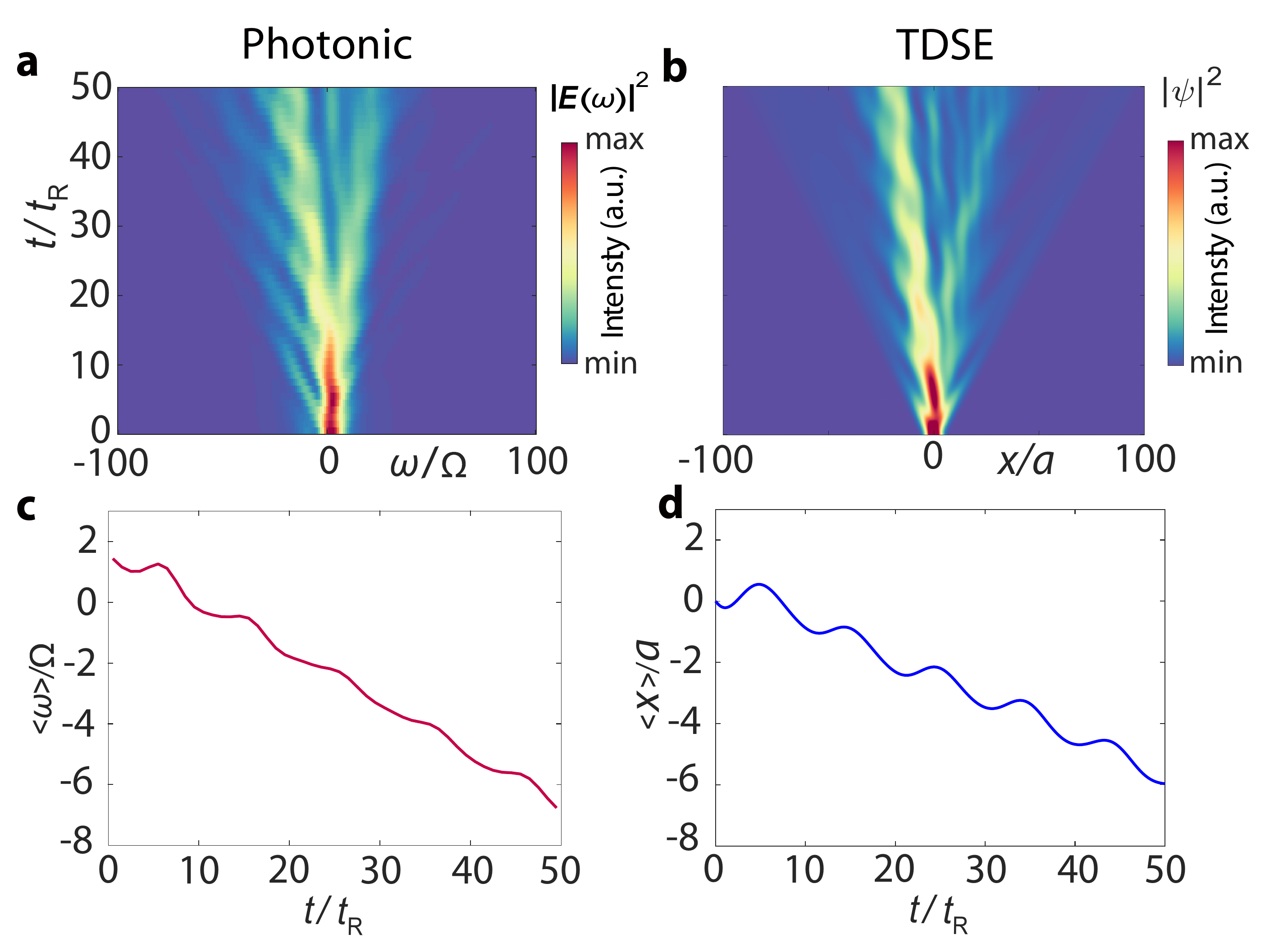}
    \centering
    \caption{\textbf{Dynamics in SU(3) synthetic gauge fields.} \textbf{a-b.} Computational photonic simulation of SU(3) dynamics and theoretical calculation of intensity by the time-dependent Schrödinger equation (TDSE). The gauge-field parameters chosen are $\theta_1 = 0.4\,\mathrm{rad}, \theta_2 = -0.5\,\mathrm{rad}\,\,\text{and}\,\,\theta_8 = 0.7\,\mathrm{rad}$, which is same as above. We also set $\omega_7 = 0.5\,\mathrm{rad\,s^{-1}}$ and $J=-1$. 
    \textbf{c-d.} Position expectation (Center of Mass C.O.M.) of the evolution by simulation and theoretical calculation, respectively. The results of simulation and theory generally agree, demonstrating the feature of Zitterbewegung-like oscillation.}
    \label{fig:SU3dynamics}
\end{figure}

\emph{Dynamics analysis.} Finally, we investigate the wave-packet dynamics induced by non-Abelian SU(3) gauge fields. Due to the complexity of the system, a full-wave photonic solution cannot be analytically obtained. Instead, we perform numerical simulations by exciting the simulated system with a single Gaussian pulse and observing its spectral evolution inside the ring after several round trips. The validity of the result from numerical simulation can be confirmed by solving the time-dependent Schrödinger equation (TDSE). The wave function can be computed by the following formula:
\begin{align}
\begin{aligned}
\psi(x,t) & =\frac{1}{\sqrt{2\pi}} \times\\
&\int_{-\infty}^\infty \diff k\, \mathbf{U}^\dagger \left(\sum_{j=0,1,2} \expn^{-\iu \tilde{E}_j t} \prod_{m \neq j} \frac{\tilde{H}(k) - \tilde{E}_m(k) I}{\tilde{E}_j(k) - \tilde{E}_m (k)}    \right)\mathbf{U} \expn^{\iu kx} \langle k | \psi(0)\rangle\,,
\end{aligned}
\end{align}
where $\langle k | \psi(0)\rangle$ is the initial state evaluated at $t=0$ and we input a Gaussian wave packet $\psi(k,0) =\langle k | \psi(0)\rangle =(1/\sqrt{2}) (2\gamma/\pi)^{1/4}e^{-\gamma k^2}(1,0,1)^T$. $\tilde{H}(k)$ is the reduced partial $\mathrm{SU(3)}$ Hamiltonian, $\{\tilde{E}_0(k),\tilde{E}_1(k),\tilde{E}_2(k)\}$ are the three non-degenerate eigen-energies of $\tilde{H}(k)$ and $\mathbf{U} = \mathrm{diag\Big((\theta_1+\iu\theta_2)/\sqrt{\theta_1^2 + \theta_2^2}\, , 1 \,,\iu \Big)}$ is introduced to simplify the calculation such that the reduced partial $\mathrm{SU(3)}$ Hamiltonian is obtained from the similarity transformation of the original partial $\mathrm{SU(3)}$ Hamiltonian with three generators $\lambda_1,\lambda_2$ and $\lambda_8$ only, i,e. $ \tilde{H}(k) : = \mathbf{U}H(k) \mathbf{U}^\dagger$. The total intensity is given by $I_{\mathrm{tot}} = |\psi|^2 = |\psi_{\mathrm{r}}(x,t)|^2 + |\psi_{\mathrm{g}}(x,t)|^2 + |\psi_{\mathrm{b}}(x,t)|^2$ and the position expectation is given by $\langle x(t)\rangle = \int_{-\infty}^{\infty} xI_{\mathrm{tot}} \diff x$. The details of the derivation are shown in SM Sec. 7. Our results show that under partial $\mathrm{SU(3)}$ group structure with color-orbit coupling given by the coupler $\mathbf{T}_{\mathrm{coc}}=\expn^{\iu \tilde{\phi} \lambda_7}$, the particles undergo an undirectional Zitterbewegung (ZB)-like dynamics, which shows a similar analog to the $\mathrm{SU(2)}$ Zitterbewegung in the previous study \cite{wong2025synthetic}. 
Fig.~\ref{fig:SU3dynamics} compares the photonic and quantum mechanical results of the evolution of the wave function, where decent agreement is achieved between them.
In the SM Sec. 7, we also investigate the pure dynamics induced by the $\mathrm{SU(3)}$ Hamiltonian when turning off the color-orbit coupling, i.e. $\omega_7=0$, and gradually increasing $\omega_y$ to $0.2\,\mathrm{rad\,s^{-1}}$, in which the ZB-like trembling micromotion begins to emerge.

\emph{Conclusions and outlook.} 
In summary, we theoretically propose a scheme to realize the $\mathrm{SU(3)}$ gauge fields using three coupled ring resonators in photonics. 
Our setup could be generalized for $\mathrm{SU}(N)$ gauge groups. The $\mathrm{SU}(N)$ gauge group can be decomposed into a product of factors of exponentials, similar to the case of  $D(\alpha,\beta,\gamma,\theta,a,b,c,\phi)$ in $\mathrm{SU(3)}$ we demonstrated above. The factorization scheme of $\mathrm{SU}(N)$ follows the reference in \cite{byrd1998differential}. Similarly, we can decompose the product of exponentials into a set of optical modulators; hence, the $\mathrm{SU}(N)$ gauge group or its subgroup can in principle also be synthesized in photonic ring resonator systems. 
%
% \remove{In particular, the case for $N=5$ is a possible hypothetical scheme for unification of $\mathrm{SU(3)\times SU(2)\times U(1)}$ for all the fundamental forces. The only experimental concern is for larger $N$, the number of optical modulators increases such that the transmission loss would be huge, of which we need to introduce more Erbium-Doped Fiber Amplifiers (EDFA) in the ring, which makes actual realization challenging. Another challenge would be the synchronization of the large number of optical modulators, for which all the optical modulators have to be in phase, would be a difficult task as $N$ grows. 
% }

\emph{Acknowledgments}. We thank Jinbing Hu for helpful discussions.  
%
%The authors acknowledge the support from the National Natural Science Foundation of China Excellent Young Scientists Fund (12222417), the Hong Kong Research Grants Council through Early Career Scheme (27300924),  Strategic Topics (STG3/E-704/23-N), the Startup Fund of The University of Hong Kong, Ms.~Belinda Hung, the Asian Young Scientist Fellowship, the Croucher Foundation, and New Cornerstone Science Foundation through the Xplorer Prize. 

%\bibliographystyle{apsrev4-1}
%\bibliography{sources0}

%

\end{document}

%-----TITLE-----
\title{
Supplementary Materials\\
Synthetic SU(3) non-Abelian gauge fields in photonic synthetic frequency dimensions}

%-----AUTHORS AND AFFILIATIONS-----
\author{Bengy Tsz Tsun Wong}
\affiliation{\hkuaffil}
\affiliation{\hkustate}
\author{Shu Yang}
\affiliation{\hkuaffil}
\affiliation{\hkustate}
\author{Zehai Pang}
\affiliation{\hkuaffil}
\affiliation{\hkustate}
\author{Yi~Yang}
\email{yiyg@hku.hk}
\affiliation{\hkuaffil}
\affiliation{\hkustate}

\maketitle

%-----CHANGE SETUP FOR PARAGRAPH INDENTS AND SKIPS-----
\setlength{\parindent}{0em}
\setlength{\parskip}{.5em}

%----- TABLE OF CONTENTS -----
\noindent{\textbf{\textsf{CONTENTS}}}\\ 
\twocolumngrid
\begingroup 
    \let\bfseries\relax % ToC sections are set in bold in RMP - we don't want that, so we relax \bfseries to nothing
    \deactivateaddvspace % removes excessive vertical spacing between ToC section lines
    \deactivatetocsubsections % removes subsections from ToC
    \tableofcontents
\endgroup
\onecolumngrid

% ----- Manual hyphenation rules -----
\hyphenation{eigen-index}

\section{Hamiltonian derivation of $\mathrm{SU(3)}$ gauge group} \label{HamSU3}
In this section, we derive the required Hamiltonian for synthesizing the full $\mathrm{SU(3)}$ gauge group in the photonic synthetic dimension. We begin with the transfer matrix of the full $\mathrm{SU(3)}$ gauge group for the modulation, which takes the following form:
\begin{equation} \label{eq:transferSU3}
\mathbf{T} =\expn^{-\iu A\cos(\Omega t + \sum_{j=1}^{\mathrm{dim\,SU(3)}}\theta^a X^a)} = \expn^{-\iu A \cos(\Omega t + \theta_1 \lambda_1 +\cdots + \theta_8 \lambda_8)}\,,
\end{equation}
where $\{\lambda_{j} \}_{j=1}^8$ are the eight Gell-Mann matrices.
According to Ref.~\cite{kaiser2022solving}, the group element of SU(3) can be written explicitly as
\begin{equation}
\expn^{\iu (\theta_1 \lambda_1 +\cdots + \theta_8 \lambda_8)}= \sum_{j=1}^3 \frac{\expn^{\iu z_j \|\theta\|}}{3z_j^2 -1} [(z_j^2 -1)I +z_j\hat{\theta}\cdot \vec{\lambda} + (\hat{\theta}\cdot \vec{\lambda})^2  ]\,,
\end{equation}
where $\|\theta\|=\sqrt{\theta_1^2 + \cdots + \theta_8^2}$, and $\hat{\theta}=\vec{\theta}/\|\theta\|$ and
\begin{equation}
z_1 = \frac{2}{\sqrt{3}} \cos \psi,\,\, z_2 = -\sin\psi -\frac{1}{\sqrt{3}}\cos\psi,\,\, z_3 = \sin \psi -\frac{1}{\sqrt{3}}\cos\psi \quad\text{with}\quad \psi =\frac{1}{3} \cos^{-1} \bigg[\frac{3\sqrt{3}}{2}\det \bigg(\frac{\vec{\theta}\cdot \vec{\lambda}}{\|\theta\|}    \bigg)\bigg]\,,
\end{equation}
satisfying the constraint of $z_1 + z_2 + z_3 =0$. Next, we want to explicitly evaluate the second-order term
\begin{equation}
\begin{aligned}
(\hat{\theta}\cdot \vec{\lambda})^2 &=\bigg(\frac{\vec{\theta}\cdot \vec{\lambda}}{\|\theta\|}    \bigg)\bigg(\frac{\vec{\theta}\cdot \vec{\lambda}}{\|\theta\|}    \bigg)\\
&= \frac{1}{\|\theta\|^2}\sum_a (\theta_a \lambda_a) \sum_b (\theta_b\lambda_b) =\frac{1}{\|\theta\|^2}\sum_a \sum_b \theta_a\theta_b \lambda_a\lambda_b\\
&=\frac{1}{2\|\theta\|^2}\sum_a\sum_b\theta_a\theta_b \{\lambda_a, \lambda_b\}\\
&=\frac{1}{2\|\theta\|^2}\sum_a\sum_b \theta_a\theta_b \bigg(\frac{4}{3}\delta_{ab}I + 2\sum_c d^{abc}\lambda_c \bigg)\\
&=\frac{2}{3}I + \frac{1}{\|\theta\|^2}\sum_{abc}d^{abc}\theta_a\theta_b \lambda_c \,.
\end{aligned}
\end{equation}
Therefore, we can write the Lie-group element of SU(3) as a linearization of Gell-Mann matrices as follows
\begin{equation}
\expn^{\iu (\theta_1 \lambda_1 +\cdots + \theta_8 \lambda_8)}= \sum_{j=1}^3 \bigg[ \expn^{\iu z_j \|\theta\|} P_j I + \sum_c \bigg(  \expn^{\iu z_j \|\theta\|} Q_j \frac{\theta_c}{\|\theta\|} + \sum_{ab} \expn^{\iu z_j \|\theta\|}R_j d^{abc}\frac{\theta_a\theta_b}{\|\theta\|^2}\bigg) \lambda_c \bigg]\,,
\end{equation}
where we have defined
\begin{equation}
P_j = \frac{z_j^2 -1/3}{3z_j^2 -1} = \frac{1}{3}\,,\quad Q_j= \frac{z_j}{3z_j^2-1}\,\quad \text{and}\,\quad R_j = \frac{1}{3z_j^2 -1}\,.
\end{equation}
For notational simplicity, from now on we will drop the repeated summation indices for $a,b$ and $c$ as understood with Einstein's summation convention. From the standard literature, the non-zero completely symmetric structural constants $d^{abc}$ are given by
\begin{equation}
\begin{cases}
& d_{118} = d_{228} = d_{338} = -d_{888} = 
\frac{1}{\sqrt{3}}\\
& d_{448} = d_{558} = d_{668} = d_{778} =-\frac{1}{2\sqrt{3}} \\
& d_{344}=d_{355} =-d_{366} = -d_{377}=-d_{247} = d_{146} =d_{157}= d_{256} = \frac{1}{2}
\end{cases} \,.
\end{equation}
With these values, we can calculate the non-zero contribution that arises from the second-order term as follows:
\begin{equation}\label{eq:166}
\begin{cases}
d^{ab1}\theta_a\theta_b\lambda_1 &= \bigg(\frac{2}{\sqrt{3}}\theta_1\theta_8 + \theta_5\theta_7\bigg)\lambda_1 \\
d^{ab2}\theta_a\theta_b\lambda_2 &= \bigg(\frac{2}{\sqrt{3}}\theta_2\theta_8 - \theta_4\theta_7 + \theta_5\theta_6\bigg)\lambda_2 \\
d^{ab3}\theta_a\theta_b\lambda_3&=\bigg(\frac{2}{\sqrt{3}}\theta_3\theta_8 +\frac{1}{2}\theta_4^2 + \frac{1}{2}\theta_5^2 -\frac{1}{2}\theta_6^2 -\frac{1}{2}\theta_7^2  \bigg)\lambda_3\\
d^{ab4}\theta_a\theta_b\lambda_4 &= \bigg(-\frac{1}{\sqrt{3}}\theta_4\theta_8 + \theta_3\theta_4 -\theta_2\theta_7 + \theta_1\theta_6 \bigg)\lambda_4 \\
d^{ab5}\theta_a\theta_b\lambda_5&= \bigg(-\frac{1}{\sqrt{3}}\theta_5\theta_8 + \theta_3\theta_5 +\theta_1\theta_7 + \theta_2\theta_6 \bigg)\lambda_5 \\
d^{ab6}\theta_a\theta_b\lambda_6 &= \bigg(-\frac{1}{\sqrt{3}}\theta_6\theta_8 - \theta_3\theta_6 +\theta_1\theta_4 + \theta_2\theta_5 \bigg)\lambda_6 \\
d^{ab7}\theta_a\theta_b\lambda_7 &= \bigg(-\frac{1}{\sqrt{3}}\theta_7\theta_8 - \theta_3\theta_7 -\theta_2\theta_4 + \theta_1\theta_5 \bigg)\lambda_7 \\
d^{ab8}\theta_a\theta_b\lambda_8 &=\bigg( \frac{1}{\sqrt{3}}(\theta_1^2 + \theta_2^2 +\theta_3^2 -\theta_8^2  ) -\frac{1}{2\sqrt{3}}(\theta_4^2 + \theta_5^2 + \theta_6^2 + \theta_7^2 ) \bigg)\lambda_8 \\
\end{cases} \,.
\end{equation}
Now, referring back to the transfer matrix, we would have
\begin{equation}
\begin{aligned}
\mathbf{T} & = \expn^{-\iu A \cos(\Omega t + \theta_1 \lambda_1 +\cdots + \theta_8 \lambda_8)}\\
&= \exp\bigg[-\frac{\iu A}{2}\bigg(\expn^{\iu (\Omega t + \theta_1 \lambda_1 +\cdots + \theta_8 \lambda_8)} + \expn^{-\iu (\Omega t + \theta_1 \lambda_1 +\cdots + \theta_8 \lambda_8)} \bigg)  \bigg] \\
&=\exp\bigg[-\iu A\sum_{j=1}^3 \bigg(P_j\cos(\Omega t + z_j \|\theta\|) \bigg)  +\sum_{j=1}^3 \cos(\Omega t + z_j \|\theta\|)\bigg(Q_j \frac{\theta_c}{\|\theta\|} + R_j d^{abc}\frac{\theta_a\theta_b}{\|\theta\|^2} \bigg)\lambda_c  \bigg] \\
&=\exp\bigg[-\iu A\sum_{j=1}^3 \bigg(P_j\cos(\Omega t + z_j \|\theta\|)\bigg)  \bigg]\exp\bigg[ -\iu A\sum_{j=1}^3 \cos(\Omega t + z_j \|\theta\|)\bigg(Q_j \frac{\theta_c}{\|\theta\|} + R_j d^{abc}\frac{\theta_a\theta_b}{\|\theta\|^2} \bigg)\lambda_c \bigg] \\
&=\left[\prod_{j=1}^3 \expn^{-\iu AP_j\cos(\Omega t + z_j \|\theta\|)}\right] \left[\prod_{j=1}^3 \expn^{-\iu A\cos(\Omega t + z_j \|\theta\|)\big(Q_j \frac{\theta_c}{\|\theta\|} + R_j d^{abc}\frac{\theta_a\theta_b}{\|\theta\|^2} \big)\lambda_c} \right]\,,
\end{aligned}
\end{equation}
where, in the last line, we have used the Zassenhaus formula to show that terms involving $\mathcal{O}(A^{n \geq 2})$ all vanish for small $A$ up to first order. By the Zassenhaus formula:
\begin{align}
\expn^{t(X+Y)} =\expn^{tX}\expn^{tY}\expn^{-\frac{t^2}{2!}[X,Y]} \expn^{\frac{t^3}{3!}(2[Y,[X,Y]]+ [X,[X,Y]])}\expn^{-\frac{t^4}{4!}([[[X,Y],X],X] + 3[[[X,Y],X],Y] +3[[[X,Y],Y],Y])}\cdots\,\,\,,
\end{align}
and we need to extend this formula with three operators $X,Y$ and $Z$. Consider a replacement of variables, $Y \rightarrow Y +Z$, this gives
\begin{equation}
\expn^{t(X+Y+Z)} = \expn^{tX}\expn^{tY}\expn^{tZ}\expn^{-\frac{t^2}{2!}[Y,Z]}\expn^{-\frac{t^2}{2!}([X,Y] + [X,Z])}\cdots \,,  
\end{equation}
here we just need to consider the non-$\mathrm{U(1)}$ contribution. For notational simplicity, we define
\begin{equation}
\tilde{\Theta}_{jc}=\cos(\Omega t + z_j \|\theta\|)\bigg[Q_j \frac{\theta_c}{\|\theta\|} + R_j d^{abc}\frac{\theta_a\theta_b}{\|\theta\|^2} \bigg]\,.
\end{equation}
Take $X = \sum_c\Theta_{1c}\lambda_c$,  $Y = \sum_g\Theta_{2g}\lambda_g$ and $Z = \sum_h\Theta_{2h}\lambda_h$; we can see that indeed the commutators do not vanish. By inspecting, for example:
\begin{align}
\begin{aligned}
[Y,Z] &= \bigg[  \sum_g\tilde{\Theta}_{2g}\lambda_g , \sum_h\tilde{\Theta}_{3h}\lambda_h \bigg]\\
&= \sum_g \sum_h \tilde{\Theta}_{2g}\tilde{\Theta}_{3h}[\lambda_g , \lambda_h] \\
&= \sum_g \sum_h \tilde{\Theta}_{2g}\tilde{\Theta}_{3h}if^{ghl}\lambda_l \\
&= \frac{\iu}{2} (\sum_g \sum_h \tilde{\Theta}_{2g}\tilde{\Theta}_{3h}f^{ghl} + \tilde{\Theta}_{2g}\tilde{\Theta}_{3h}f^{ghl}\lambda_l) \\
&=\frac{\iu}{2} \sum_g \sum_h (\tilde{\Theta}_{2g}\tilde{\Theta}_{3h}f^{ghl} + \tilde{\Theta}_{2h}\tilde{\Theta}_{3g}f^{hgl})\lambda_l
\\
&=\frac{\iu}{2} \sum_g \sum_h (\tilde{\Theta}_{2g}\tilde{\Theta}_{3h}- \tilde{\Theta}_{2h}\tilde{\Theta}_{3g})f^{ghl}\lambda_l\\
&\neq 0
\end{aligned}
\end{align}
due to the fact that $\tilde{\Theta}_{2g}\tilde{\Theta}_{3h}- \tilde{\Theta}_{2h}\tilde{\Theta}_{3g} \neq 0$. However, as $A$ is small, we only keep up to the first order in $A$, so by replacing $t\rightarrow -\iu A$,
\begin{equation} \label{eq:BCHapprox}
\expn^{-\iu A(X+Y+Z)} \approx \expn^{-\iu AX}\expn^{-\iu AY}\expn^{-\iu AZ}  \,.
\end{equation}
Therefore, we finally obtain
\begin{equation}
\boxed{
\mathbf{T} = \expn^{-\iu A \cos(\Omega t + \theta_1 \lambda_1 +\cdots + \theta_8 \lambda_8)} = \left[\prod_{j=1}^3 \expn^{-\iu AP_j\cos(\Omega t + z_j \|\theta\|)}\right] \left[\prod_{j=1}^3 \expn^{-\iu A\cos(\Omega t + z_j \|\theta\|)\big(Q_j \frac{\theta_c}{\|\theta\|} + R_j d^{abc}\frac{\theta_a\theta_b}{\|\theta\|^2} \big)\lambda_c} \right]
} \,.
\end{equation}
Thus the transfer matrix is the product of $\mathrm{U(1)\times U(1) \times U(1)}$ and the product of $\mathrm{SU(3)\times SU(3) \times  SU(3)}$. Given that $A$ is small, we can Taylor expand the transfer matrix up to first order as follows:
\begin{equation}
\mathbf{T} = 1-\iu A\sum_j P_j\cos(\Omega t + z_j \|\theta\|) -\iu A\sum_j \cos(\Omega t + z_j \|\theta\|)\bigg[Q_j \frac{\theta_c}{\|\theta\|} + R_j d^{abc}\frac{\theta_a\theta_b}{\|\theta\|^2}  \bigg]\lambda_c + \mathcal{O}(A^2) \,.
\end{equation} 
The transfer matrix acts on the state of
\begin{equation}
\psi_n =\begin{pmatrix}
a_{n\rmred}\\
a_{n\rmgreen}\\
a_{n\rmblue}
\end{pmatrix}\,,
\end{equation}
where $\mathrm{r}, \mathrm{g}, \mathrm{b}$ stand for ``red'', ``green'' and ``blue'' modes, mimicking color charges in quantum chromodynamics. Acting the transfer matrix on the state and multiplying $\iu /\tR$ on both sides, and with some rearrangement, we obtain
\begin{equation}
\begingroup\makeatletter\def\f@size{7.5}\check@mathfonts
\iu\begin{pmatrix}
\frac{a_{n\rmred}^\prime - a_{n\rmred}}{\tR}\\
\frac{a_{n\rmgreen}^\prime - a_{n\rmgreen}}{\tR}\\
\frac{a_{n\rmblue}^\prime - a_{n\rmblue}}{\tR}\\
\end{pmatrix} 
=\frac{A}{\tR}
\begin{pmatrix}
\sum_{j=1}^3\big[P_j\cos(\Omega t +z_j\|\theta\|)+\big(\Theta_{j3} + \frac{\Theta_{j8}}{\sqrt{3}}  \big)\big] & \sum_{j=1}^3 (\Theta_{j1}-\iu \Theta_{j2}) & \sum_{j=1}^3 (\Theta_{j4}-\iu \Theta_{j5}) \\
\sum_{j=1}^3 (\Theta_{j1}+\iu \Theta_{j2})
& \sum_{j=1}^3\big[P_j\cos(\Omega t +z_j\|\theta\|)+\big(-\Theta_{j3} + \frac{\Theta_{j8}}{\sqrt{3}}  \big)\big] & \sum_{j=1}^3 (\Theta_{j6}-\iu \Theta_{j7}) \\
\sum_{j=1}^3 (\Theta_{j4}+\iu \Theta_{j5}) & \sum_{j=1}^3 (\Theta_{j6}+\iu \Theta_{j7}) &\sum_{j=1}^3\big[P_j\cos(\Omega t +z_j\|\theta\|)-\frac{2\Theta_{j8}}{\sqrt{3}}
\end{pmatrix}
\begin{pmatrix}
a_{n\rmred}\\
a_{n\rmgreen}\\
a_{n\rmblue}
\end{pmatrix}
\endgroup\,.
\end{equation}
Then we will have three coupled differential equations of motion as follows, firstly for the red mode
\begin{equation} \label{eq:EOM1}
\begin{aligned}
\iu \frac{\diff a_{n\rmred}}{\diff t} &= J\sum_{j=1}^3\bigg\{\expn^{\iu z_j\|\theta\|}\bigg[P_j + Q_j\frac{1}{\|\theta\|} \bigg(\theta_3 + \frac{1}{\sqrt{3}}\theta_8 \bigg)+R_j\frac{1}{\|\theta\|^2}\bigg(d^{ab3}+\frac{1}{\sqrt{3}}d^{ab8}\bigg)\theta_a\theta_b\bigg] a_{n-1\,\rmred} \\
&\quad\quad\quad\quad +\expn^{-\iu z_j\|\theta\|}\bigg[P_j + Q_j\frac{1}{\|\theta\|} \bigg(\theta_3 + \frac{1}{\sqrt{3}}\theta_8 \bigg)+R_j\frac{1}{\|\theta\|^2}\bigg(d^{ab3}+\frac{1}{\sqrt{3}}d^{ab8}\bigg)\theta_a\theta_b\bigg] a_{n+1\,\rmred}\\
&\quad\quad\quad\quad + \expn^{\iu z_j\|\theta\|}\bigg[Q_j\frac{1}{\|\theta\|}(\theta_1 -\iu\theta_2)+R_j\frac{1}{\|\theta\|^2}(d^{ab1}-\iu d^{ab2}\theta_a\theta_b)\bigg]a_{n-1\,\rmgreen}\\
&\quad\quad\quad\quad + \expn^{-\iu z_j\|\theta\|}\bigg[Q_j\frac{1}{\|\theta\|}(\theta_1 -\iu\theta_2)+R_j\frac{1}{\|\theta\|^2}(d^{ab1}-\iu d^{ab2}\theta_a\theta_b)\bigg]a_{n+1\,\rmgreen}\\
&\quad\quad\quad\quad + \expn^{\iu z_j\|\theta\|}\bigg[Q_j\frac{1}{\|\theta\|}(\theta_4 -\iu\theta_5)+R_j\frac{1}{\|\theta\|^2}(d^{ab4}-\iu d^{ab5}\theta_a\theta_b)\bigg]a_{n-1\,\rmblue}\\
&\quad\quad\quad\quad + \expn^{-\iu z_j\|\theta\|}\bigg[Q_j\frac{1}{\|\theta\|}(\theta_4 -\iu\theta_5)+R_j\frac{1}{\|\theta\|^2}(d^{ab4}-\iu d^{ab5}\theta_a\theta_b)\bigg]a_{n+1\,\rmblue}
\bigg\}\,.
\end{aligned}
\end{equation}
Secondly for the green mode,
\begin{equation} \label{eq:EOM2}
\begin{aligned}
\iu \frac{\diff a_{n\rmgreen}}{\diff t} &= J\sum_{j=1}^3\bigg\{\expn^{\iu z_j\|\theta\|}\bigg[P_j + Q_j\frac{1}{\|\theta\|} \bigg(-\theta_3 + \frac{1}{\sqrt{3}}\theta_8 \bigg)+R_j\frac{1}{\|\theta\|^2}\bigg(-d^{ab3}+\frac{1}{\sqrt{3}}d^{ab8}\bigg)\theta_a\theta_b\bigg] a_{n-1\,\rmgreen} \\
&\quad\quad\quad\quad +\expn^{-\iu z_j\|\theta\|}\bigg[P_j + Q_j\frac{1}{\|\theta\|} \bigg(-\theta_3 + \frac{1}{\sqrt{3}}\theta_8 \bigg)+R_j\frac{1}{\|\theta\|^2}\bigg(-d^{ab3}+\frac{1}{\sqrt{3}}d^{ab8}\bigg)\theta_a\theta_b\bigg] a_{n+1\,\rmgreen}\\
&\quad\quad\quad\quad + \expn^{\iu z_j\|\theta\|}\bigg[Q_j\frac{1}{\|\theta\|}(\theta_1 +\iu\theta_2)+R_j\frac{1}{\|\theta\|^2}(d^{ab1}+\iu d^{ab2}\theta_a\theta_b)\bigg]a_{n-1\,\rmred}\\
&\quad\quad\quad\quad + \expn^{-\iu z_j\|\theta\|}\bigg[Q_j\frac{1}{\|\theta\|}(\theta_1+\iu\theta_2)+R_j\frac{1}{\|\theta\|^2}(d^{ab1}+\iu d^{ab2}\theta_a\theta_b)\bigg]a_{n+1\,\rmred}\\
&\quad\quad\quad\quad + \expn^{\iu z_j\|\theta\|}\bigg[Q_j\frac{1}{\|\theta\|}(\theta_6 -\iu\theta_7)+R_j\frac{1}{\|\theta\|^2}(d^{ab6}-\iu d^{ab7}\theta_a\theta_b)\bigg]a_{n-1\,\rmblue}\\
&\quad\quad\quad\quad + \expn^{-\iu z_j\|\theta\|}\bigg[Q_j\frac{1}{\|\theta\|}(\theta_6 -\iu\theta_7)+R_j\frac{1}{\|\theta\|^2}(d^{ab6}-\iu d^{ab7}\theta_a\theta_b)\bigg]a_{n+1\,\rmblue} 
\bigg\}\,.
\end{aligned}
\end{equation}
Finally for the blue mode,
\begin{equation} \label{eq:EOM3}
\begin{aligned}
\iu \frac{\diff a_{n\rmblue}}{\diff t} &= J\sum_{j=1}^3\bigg\{\expn^{\iu z_j\|\theta\|}\bigg[P_j + Q_j\frac{1}{\|\theta\|} \bigg( \frac{-2}{\sqrt{3}}\theta_8 \bigg)+R_j\frac{1}{\|\theta\|^2}\bigg(\frac{-2}{\sqrt{3}}d^{ab8}\bigg)\theta_a\theta_b\bigg] a_{n-1\,\rmblue} \\
&\quad\quad\quad\quad +\expn^{-\iu z_j\|\theta\|}\bigg[P_j + Q_j\frac{1}{\|\theta\|} \bigg(\frac{-2}{\sqrt{3}}\theta_8 \bigg)+R_j\frac{1}{\|\theta\|^2}\bigg(\frac{-2}{\sqrt{3}}d^{ab8}\bigg)\theta_a\theta_b\bigg] a_{n+1\,\rmblue}\\
&\quad\quad\quad\quad + \expn^{\iu z_j\|\theta\|}\bigg[Q_j\frac{1}{\|\theta\|}(\theta_4 +\iu\theta_5)+R_j\frac{1}{\|\theta\|^2}(d^{ab4}+\iu d^{ab5}\theta_a\theta_b)\bigg]a_{n-1\,\rmred}\\
&\quad\quad\quad\quad + \expn^{-\iu z_j\|\theta\|}\bigg[Q_j\frac{1}{\|\theta\|}(\theta_4+\iu\theta_5)+R_j\frac{1}{\|\theta\|^2}(d^{ab4}+\iu d^{ab5}\theta_a\theta_b)\bigg]a_{n+1\,\rmred}\\
&\quad\quad\quad\quad + \expn^{\iu z_j\|\theta\|}\bigg[Q_j\frac{1}{\|\theta\|}(\theta_6 +\iu\theta_7)+R_j\frac{1}{\|\theta\|^2}(d^{ab6}+\iu d^{ab7}\theta_a\theta_b)\bigg]a_{n-1\,\rmgreen}\\
&\quad\quad\quad\quad + \expn^{-\iu z_j\|\theta\|}\bigg[Q_j\frac{1}{\|\theta\|}(\theta_6 +\iu\theta_7)+R_j\frac{1}{\|\theta\|^2}(d^{ab6}+\iu d^{ab7}\theta_a\theta_b)\bigg]a_{n+1\,\rmgreen}
\bigg\} \,.
\end{aligned}
\end{equation}
Here $J=A/2\tR$. From these equations of motion, we can recover the corresponding Hamiltonian. From \ref{eq:EOM1}, first we have
\begin{equation} 
\begin{aligned}
\hat{H}_1 &= J\sum_{j=1}^3\bigg\{\expn^{\iu z_j\|\theta\|}\bigg[P_j + Q_j\frac{1}{\|\theta\|} \bigg(\theta_3 + \frac{1}{\sqrt{3}}\theta_8 \bigg)+R_j\frac{1}{\|\theta\|^2}\bigg(d^{ab3}+\frac{1}{\sqrt{3}}d^{ab8}\bigg)\theta_a\theta_b\bigg] \hat{a}^\dagger_{n+1\,\rmred}\hat{a}_{n\,\rmred} \\
&\quad\quad\quad\quad +\expn^{-\iu z_j\|\theta\|}\bigg[P_j + Q_j\frac{1}{\|\theta\|} \bigg(\theta_3 + \frac{1}{\sqrt{3}}\theta_8 \bigg)+R_j\frac{1}{\|\theta\|^2}\bigg(d^{ab3}+\frac{1}{\sqrt{3}}d^{ab8}\bigg)\theta_a\theta_b\bigg] \hat{a}^\dagger_{n\,\rmred}\hat{a}_{n+1\,\rmred}\\
&\quad\quad\quad\quad + \expn^{\iu z_j\|\theta\|}\bigg[Q_j\frac{1}{\|\theta\|}(\theta_1 -\iu\theta_2)+R_j\frac{1}{\|\theta\|^2}(d^{ab1}-\iu d^{ab2}\theta_a\theta_b)\bigg]\hat{a}^\dagger_{n+1\,\rmred}\hat{a}_{n\,g}\\
&\quad\quad\quad\quad + \expn^{-\iu z_j\|\theta\|}\bigg[Q_j\frac{1}{\|\theta\|}(\theta_1 -\iu\theta_2)+R_j\frac{1}{\|\theta\|^2}(d^{ab1}-\iu d^{ab2}\theta_a\theta_b)\bigg]\hat{a}^\dagger_{n\,\rmred}\hat{a}_{n+1\,g}\\
&\quad\quad\quad\quad + \expn^{\iu z_j\|\theta\|}\bigg[Q_j\frac{1}{\|\theta\|}(\theta_4 -\iu\theta_5)+R_j\frac{1}{\|\theta\|^2}(d^{ab4}-\iu d^{ab5}\theta_a\theta_b)\bigg]\hat{a}^\dagger_{n+1\,\rmred}\hat{a}_{n\,\rmblue}\\
&\quad\quad\quad\quad + \expn^{-\iu z_j\|\theta\|}\bigg[Q_j\frac{1}{\|\theta\|}(\theta_4 -\iu\theta_5)+R_j\frac{1}{\|\theta\|^2}(d^{ab4}-\iu d^{ab5}\theta_a\theta_b)\bigg]\hat{a}^\dagger_{n\,\rmred}\hat{a}_{n+1\,\rmblue}
\bigg\}\,.
\end{aligned}
\end{equation}
Next, from~\eqref{eq:EOM2} we have
\begin{equation} 
\begin{aligned}
\hat{H}_2 &= J\sum_{j=1}^3\bigg\{\expn^{\iu z_j\|\theta\|}\bigg[P_j + Q_j\frac{1}{\|\theta\|} \bigg(-\theta_3 + \frac{1}{\sqrt{3}}\theta_8 \bigg)+R_j\frac{1}{\|\theta\|^2}\bigg(-d^{ab3}+\frac{1}{\sqrt{3}}d^{ab8}\bigg)\theta_a\theta_b\bigg]  \hat{a}^\dagger_{n+1\,g}\hat{a}_{n\,g} \\
&\quad\quad\quad\quad +\expn^{-\iu z_j\|\theta\|}\bigg[P_j + Q_j\frac{1}{\|\theta\|} \bigg(-\theta_3 + \frac{1}{\sqrt{3}}\theta_8 \bigg)+R_j\frac{1}{\|\theta\|^2}\bigg(-d^{ab3}+\frac{1}{\sqrt{3}}d^{ab8}\bigg)\theta_a\theta_b\bigg]\hat{a}^\dagger_{n\,g}\hat{a}_{n+1\,g}\\
&\quad\quad\quad\quad + \expn^{\iu z_j\|\theta\|}\bigg[Q_j\frac{1}{\|\theta\|}(\theta_1 +\iu\theta_2)+R_j\frac{1}{\|\theta\|^2}(d^{ab1}+\iu d^{ab2}\theta_a\theta_b)\bigg]\hat{a}^\dagger_{n+1\,g}\hat{a}_{n\,\rmred}\\
&\quad\quad\quad\quad + \expn^{-\iu z_j\|\theta\|}\bigg[Q_j\frac{1}{\|\theta\|}(\theta_1+\iu\theta_2)+R_j\frac{1}{\|\theta\|^2}(d^{ab1}+\iu d^{ab2}\theta_a\theta_b)\bigg]\hat{a}^\dagger_{n\,g}\hat{a}_{n+1\,\rmred}\\
&\quad\quad\quad\quad + \expn^{\iu z_j\|\theta\|}\bigg[Q_j\frac{1}{\|\theta\|}(\theta_6 -\iu\theta_7)+R_j\frac{1}{\|\theta\|^2}(d^{ab6}-\iu d^{ab7}\theta_a\theta_b)\bigg]\hat{a}^\dagger_{n+1\,g}\hat{a}_{n\,\rmblue}\\
&\quad\quad\quad\quad + \expn^{-\iu z_j\|\theta\|}\bigg[Q_j\frac{1}{\|\theta\|}(\theta_6 -\iu\theta_7)+R_j\frac{1}{\|\theta\|^2}(d^{ab6}-\iu d^{ab7}\theta_a\theta_b)\bigg]\hat{a}^\dagger_{n\,g}\hat{a}_{n+1\,\rmblue}
\bigg\}\,.
\end{aligned}
\end{equation}
Finally, from~\eqref{eq:EOM3} we have
\begin{equation} 
\begin{aligned}
\hat{H}_3 &= J\sum_{j=1}^3\bigg\{\expn^{\iu z_j\|\theta\|}\bigg[P_j + Q_j\frac{1}{\|\theta\|} \bigg( \frac{-2}{\sqrt{3}}\theta_8 \bigg)+R_j\frac{1}{\|\theta\|^2}\bigg(\frac{-2}{\sqrt{3}}d^{ab8}\bigg)\theta_a\theta_b\bigg] \hat{a}^\dagger_{n+1\,\rmblue}\hat{b}_{n\,g}  \\
&\quad\quad\quad\quad +\expn^{-\iu z_j\|\theta\|}\bigg[P_j + Q_j\frac{1}{\|\theta\|} \bigg(\frac{-2}{\sqrt{3}}\theta_8 \bigg)+R_j\frac{1}{\|\theta\|^2}\bigg(\frac{-2}{\sqrt{3}}d^{ab8}\bigg)\theta_a\theta_b\bigg] \hat{a}^\dagger_{n\,\rmblue}\hat{a}_{n+1\,\rmblue}\\
&\quad\quad\quad\quad + \expn^{\iu z_j\|\theta\|}\bigg[Q_j\frac{1}{\|\theta\|}(\theta_4 +\iu\theta_5)+R_j\frac{1}{\|\theta\|^2}(d^{ab4}+\iu d^{ab5}\theta_a\theta_b)\bigg]\hat{a}^\dagger_{n+1\,\rmblue}\hat{a}_{n\,\rmred}\\
&\quad\quad\quad\quad + \expn^{-\iu z_j\|\theta\|}\bigg[Q_j\frac{1}{\|\theta\|}(\theta_4+\iu\theta_5)+R_j\frac{1}{\|\theta\|^2}(d^{ab4}+\iu d^{ab5}\theta_a\theta_b)\bigg]\hat{a}^\dagger_{n\,\rmblue}\hat{a}_{n+1\,\rmred}\\
&\quad\quad\quad\quad + \expn^{\iu z_j\|\theta\|}\bigg[Q_j\frac{1}{\|\theta\|}(\theta_6 +\iu\theta_7)+R_j\frac{1}{\|\theta\|^2}(d^{ab6}+\iu d^{ab7}\theta_a\theta_b)\bigg]\hat{a}^\dagger_{n+1\,\rmblue}\hat{a}_{n\,g}\\
&\quad\quad\quad\quad + \expn^{-\iu z_j\|\theta\|}\bigg[Q_j\frac{1}{\|\theta\|}(\theta_6 +\iu\theta_7)+R_j\frac{1}{\|\theta\|^2}(d^{ab6}+\iu d^{ab7}\theta_a\theta_b)\bigg]\hat{a}^\dagger_{n\,\rmblue}\hat{a}_{n+1\,g}
\bigg\}\,.
\end{aligned}
\end{equation}
Therefore, the full Hamiltonian is given by $\hat{H}_0 = \hat{H}_1 + \hat{H}_2 + \hat{H}_3$. It can be checked that, this is equivalent to the tight-binding Hamiltonian with synthetic $\mathrm{SU(3)}$ gauge field
\begin{equation} \label{eq:HamiltonianSU3}
\boxed{
\hat{H}_0 = J \sum_{n}\sum_{\alpha,\,\beta \in\{\rmred, \rmgreen, \rmblue \}} \bigg(\hat{a}^\dagger_{n+1\,\alpha}[\expn^{i(\theta_1\lambda_1 +\cdots+\theta_8 \lambda_8)}]_{\alpha\beta}\hat{a}_{n\beta} + \hat{a}^\dagger_{n\,\alpha}[\expn^{-i(\theta_1\lambda_1 +\cdots+\theta_8 \lambda_8)}]_{\alpha\beta}\hat{a}_{n+1\,\beta} \bigg)} \,.
\end{equation}
In momentum space, the Hamiltonian reads:
\begin{equation}
\hat{H}_0 (k) = \sum_k \sum_{\alpha,\,\beta \in\{\rmred, \rmgreen, \rmblue \}} 2J \bigg[\cos(k +\theta_1\lambda_1 +\cdots + \theta_8\lambda
_8)  \bigg]_{\alpha\beta}\hat{a}^\dagger_{k\alpha}\hat{a}_{k\beta}  \,.
\end{equation}
Explicitly,
\begin{equation}
\boxed{
\hat{H}_0 (k) =2J \sum_{j=1}^3\sum_k  \sum_{\alpha,\,\beta \in\{\rmred, \rmgreen, \rmblue \}} \bigg[P_j\cos(k +z_j \|\theta\|)I +\cos(k + z_j \|\theta\|)\bigg(Q_j \frac{\theta_c}{\|\theta\|} +R_j d^{abc}\frac{\theta_a \theta_b}{\|\theta\|^2}  \bigg)\lambda_c \bigg]_{\alpha\beta}\hat{a}_{k\alpha}^\dagger \hat{a}_{k\beta}} \,.
\end{equation}

\section{Full SU(3) Hamiltonian}
In this section, we provide an analytical expression of the full SU(3) Hamiltonian. The Bloch Hamiltonian of the full SU(3) case is given by the following:
\begin{equation} \label{eq:FULLHAM}
    H_0(k) = 2J \cos(k+ \theta_1\lambda_1 + \theta_2\lambda_2+\cdots+ \theta_8\lambda_8) = J \left(\expn^{\iu(k+\theta_1\lambda_1 + \theta_2\lambda_2+\cdots+ \theta_8\lambda_8)} + \expn^{-\iu(k+\theta_1\lambda_1 + \theta_2\lambda_2+\cdots+ \theta_8\lambda_8)} \right)\,.
\end{equation}
We define the SU(3) Lie vector as $X = \sum_{j=1}^8 \theta_j \lambda_j$ and the modulus $\| \theta\| = \sqrt{\theta_1^2 + \theta_2^2 + \cdots + \theta_8^2}$. Let $\mu_{j}$ for $j=1,2,3$ be the three eigenvalues of $X$. It can be shown that the eigenvalues are
\begin{equation}
    \mu_j = \frac{2}{\sqrt{3}} \| \theta\| \cos\left(\psi + \frac{2\pi(j-1)}{3} \right) \,\,\text{for}\,j=1,2,3 \,\,\,\text{with} \,\,\, \psi = \frac{1}{3}\cos^{-1}\left(\frac{3 \sqrt{3}}{2} \frac{\mathrm{det}\,X}{\| \theta\|^3} \right)\,,
\end{equation}
where the full expression of the determinant of $\eta$ is given by
\begin{equation}
\begin{aligned}
    \mathrm{det}\,X &= \frac{2}{\sqrt{3}} \theta_8 (\theta_1^2 + \theta_2^2 + \theta_3^2) + 2\theta_1 (\theta_4 \theta_6 + \theta_5 \theta_7) + 2\theta_2 (\theta_5 \theta_6 -\theta_4 \theta_7) + \theta_3 (\theta_4^2 +\theta_5^2 -\theta_6^2 - \theta_7^2)\\
    &\quad\quad -\frac{1}{\sqrt{3}}\theta_8 (\theta_4^2 + \theta_5^2 + \theta_6^2 +\theta_7^2) -\frac{2}{3\sqrt{3}}\theta_8^3\\
    &= \frac{2}{3}\sum_{abc}d_{abc}\theta_a\theta_b \theta_c\,.
\end{aligned}
\end{equation}
To proceed, in order to obtain an analytical expression, we need to employ Sylvester's theorem. Let $\mathbf{A}$ be a square matrix of any rank and $f(\mathbf{A})$ is an analytic function of  $\mathbf{A}$, the Sylverster's theorem states that
\begin{equation}
    f(\mathbf{A}) = \sum_{i=1}^n f(\xi_i) \mathbf{P}_i \,,
\end{equation}
where $\{\xi_i\}_{i=1}^n$ are the non-degenerate eigenvalues of $\mathbf{A}$ and
\begin{equation}
    \mathbf{P}_i = \prod_{j=1,\,j\neq i}^n \frac{1}{\xi_i-\xi_j}(\mathbf{A}- \xi_j I)\,.
\end{equation}
are the corresponding Frobenius covariants of $\mathbf{A}$. In full, we have
\begin{equation}
     f(\mathbf{A}) = \sum_{i=1}^n f(\xi_i) \prod_{j=1,\,j\neq i}^n \frac{1}{\xi_i-\xi_j}(\mathbf{A}- \xi_j I)\,.
\end{equation}
Now, returning to our case of the partial $\mathrm{SU(3)}$ Hamiltonian, let $\mathbf{A} = X$ and the map $f$ be the exponential map, it follows that
\begin{equation}
    \expn^{\iu \eta } = \sum_{j=1}^3 \expn^{\iu \mu_j } \prod_{m \neq j} \frac{X - \mu_m I}{\mu_j- \mu_m}\,.
\end{equation}
Therefore, by using Sylvester’s theorem, the Bloch Hamiltonian of the full SU(3) group can be reformulated as
\begin{equation}
    \begin{aligned}
        H_0(k) &= J \left[\expn^{\iu k}\sum_{j=1}^3 \expn^{\iu \mu_j} \left(\prod_{m\neq j} \frac{X-\mu_m I}{\mu_j - \mu_m} \right) + \expn^{-\iu k}\sum_{j=1}^3 \expn^{-\iu \mu_j} \left(\prod_{m\neq j} \frac{X-\mu_m I}{\mu_j - \mu_m} \right)\right] \\
        &= 2J\cos(k+\mu_1)\frac{(X -\mu_2I)(X -\mu_3I)}{(\mu_1 -\mu_2)(\mu_1 - \mu_3)} + 2J\cos(k+\mu_2)\frac{(X -\mu_1I)(X -\mu_3I)}{(\mu_2 -\mu_1)(\mu_2 - \mu_3)} + 2J\cos(k+\mu_3)\frac{(X -\mu_1I)(X -\mu_2I)}{(\mu_3 -\mu_1)(\mu_3 - \mu_2)}\,.
    \end{aligned}
\end{equation}
We can compute the denominator as follows:
\begin{equation}
    \mu_1 - \mu_2 = 2 \| \theta\| \sin\left(\psi + \frac{\pi}{3}\right)\,\,,\,\, \mu_1 - \mu_3 = 2 \| \theta\| \sin\left(\psi + \frac{2\pi}{3}\right) \,\, \text{and} \,\, \mu_2 - \mu_3 = -2 \| \theta\| \sin\psi\,.
\end{equation}
For simplicity, we further define the $k-$dependent coefficient of each matrix as
\begin{equation}
 A(k) = \frac{J\cos(k + \mu_1)}{2\| \theta\|^2\sin\left( \psi + \frac{\pi}{3}\right) \sin\left( \psi + \frac{2\pi}{3}\right)} \,\,,\,\,B(k) =\frac{J\cos(k + \mu_2)}{2\| \theta\|^2\sin\left( \psi + \frac{\pi}{3}\right) \sin\psi}\,\,\text{and}\,\, C(k) =-\frac{J\cos(k + \mu_3)}{2\|\theta\|^2 \sin\psi\sin\left( \psi + \frac{2\pi}{3}\right) }\,.
\end{equation}
Then the Hamiltonian can be calculated as
\begin{equation}
  H_0(k) = (A+B+C)X^2 + (A\mu_1 + B\mu_2 +C \mu_3)X + (A\mu_2 \mu_3 + B\mu_1\mu_3 + C\mu_1\mu_2) I\,.  
\end{equation}

Next, we can compute the explicit matrix elements of $H(k)$. These results are given by:
\begin{equation}
    \begin{cases}
    (H_0)_{11} &= (A+B+C) \bigg[\theta_1^2 + \theta_2^2 + \theta_4^2 + \theta_5^2 +\bigg(\theta_3+\frac{\theta_8^2}{\sqrt{3}}\bigg)^2\bigg] +(A\mu_1+B\mu_2 +C\mu_3)\bigg(\theta_3+\frac{\theta_8^2}{\sqrt{3}}\bigg) + (A\mu_2 \mu_3 + B\mu_1\mu_3 + C\mu_1\mu_2)\\
    (H_0)_{22} &= (A+B+C)\bigg[\theta_1^2 + \theta_2^2 +\theta_6^2 +\theta_7^2 + \left(-\theta_3 + \frac{\theta_8}{\sqrt
    {3}} \right)^2 \bigg] +(A\mu_1+B\mu_2 +C\mu_3)\bigg(-\theta_3+\frac{\theta_8^2}{\sqrt{3}}\bigg) + (A\mu_2 \mu_3 + B\mu_1\mu_3 + C\mu_1\mu_2) \\
    (H_0)_{33} &= (A+B+C)\bigg( \theta_4^2 + \theta_5^2 + \theta_6^2 +\theta_7^2 + \frac{4}{3}\theta_8^2 \bigg) -\frac{2\theta_8}{\sqrt{3}}(A\mu_1+B\mu_2 +C\mu_3) + (A\mu_2 \mu_3 + B\mu_1\mu_3 + C\mu_1\mu_2)\\
    (H_0)_{12} &= (\theta_1 -\iu \theta_2)\left[ (A+B+C)\frac{2\theta_8}{\sqrt{3}} + (A\mu_1 + B\mu_2 +C\mu_3)\right] + (A+B+C) (\theta_4 -\iu\theta_5)(\theta_6+i\theta_7)\\
    (H_0)_{13} &= (\theta_4 -\iu \theta_5)\left[ (A+B+C)\left(\theta_3-\frac{\theta_8}{\sqrt{3}}\right) + (A\mu_1 + B\mu_2 +C\mu_3)\right] + (A+B+C) (\theta_1 -\iu\theta_2)(\theta_6-i\theta_7)\\
    (H_0)_{23} &= (\theta_6 -\iu \theta_7)\left[ (A+B+C)\left(-\theta_3-\frac{\theta_8}{\sqrt{3}}\right) + (A\mu_1 + B\mu_2 +C\mu_3)\right] + (A+B+C) (\theta_1 + \iu\theta_2)(\theta_4-i\theta_5)\\
    \end{cases}
\end{equation}
and $H_{21} = H^{*}_{12}$, $H_{31} = H^*_{13}$ and $H_{32} = H_{23}^*$.

\section{Photonic setup for full $\mathrm{SU(3)}$ gauge group}
In this section, we would investigate how to realize the full $\mathrm{SU(3)}$ gauge group in photonic synthetic dimension. From the full Hamiltonian in Eq.~ \eqref{eq:FULLHAM} , since $kI$ commutes with $X$, therefore $\expn^{\pm \iu (kI + X)} = \expn^{\pm \iu k} \expn^{\pm \iu X}$. Also, as it follows that $\expn^{\pm \iu X} = U \mathrm{diag} (\expn^{\pm \iu\mu_1} , \expn^{\pm \iu \mu_2} , \expn^{\pm \iu\mu_3}   )    U^\dagger$. Thus the Hamiltonian can be diagonalized by $U$ and $U^\dagger $ as follows:
\begin{equation}
\tilde{H}_0(k) = U^\dagger H_0(k) U = \begin{pmatrix}
 2J\cos(k+\mu_1) & 0 & 0\\
 0 &  2J\cos(k+\mu_2) & 0 \\
 0   &   0  & 2J\cos(k+\mu_3)\\
\end{pmatrix} \,.
\end{equation}
It is important to note that we can use the same $U$ and $U^\dagger$ to diagonalize both $X$ and $H(k)$, where $ U$ and $U^\dagger$ are independent of $k$.

Now, we compute the Lie vector $X$ explicitly, given by
\begin{equation}
X=\sum_{a=1}^8 \theta_a \lambda_a = \begin{pmatrix}
\theta_3+\frac{\theta_8}{\sqrt{3}} & \theta_1 - i\theta_2 & \theta_4 -i\theta_5 \\
\theta_1 +i\theta_2 & -\theta_3 + \frac{\theta_8}{\sqrt{3}} &\theta_6 -i\theta_7\\
\theta_4+i\theta_5 & \theta_6+i\theta_7 & \frac{-2\theta_8}{\sqrt{3}}\\
\end{pmatrix} \equiv
\begin{pmatrix}
a & u & v\\
u^* & b & w\\
v^* & w^* &c
\end{pmatrix}\,.
\end{equation}
The eigenvectors of the Lie vector $X$ are found to be
\begin{equation}
\tilde{v}_j(\theta_1 , \theta_2 , \cdots ,\theta_8) =\begin{pmatrix}
(b-\mu_j)(c-\mu_j) - |w|^2 \\
wv^* - u^*(c-\mu_j)\\
u^*v^* -v^*(b-\mu_j)
\end{pmatrix}\,.
\end{equation}
The normalized eigenvectors are defined by
\begin{equation}
v_j (\theta_1 , \theta_2 , \cdots ,\theta_8)= \frac{\tilde{v}_j(\theta_1 , \theta_2 , \cdots ,\theta_8)}{\| \tilde{v}_j (\theta_1 , \theta_2 , \cdots ,\theta_8)\|} \quad \text{where} \quad \| \tilde{v}_j \| =\sqrt{\tilde{v}_j^\dagger \tilde{v}_j}\,.
\end{equation}
Then the $U$ unitary matrix is given by
\begin{equation}
\boxed{
U(\theta_1 , \theta_2 , \cdots ,\theta_8) = \Big[v_1 (\theta_1 , \theta_2 , \cdots ,\theta_8) \quad v_2 (\theta_1 , \theta_2 , \cdots ,\theta_8) \quad v_3 (\theta_1 , \theta_2 , \cdots ,\theta_8) \Big] \in \mathrm{U(3)}}\,.
\end{equation}
which belongs to $\mathrm{U(3)}$. This is because the Hamiltonian $H$ is Hermitian; it is guaranteed to be diagonalized by a unitary matrix. Notice that any $\mathrm{U(3)}$ elements can be decomposed by nine Euler angles:
\begin{equation}
D(\delta, \alpha,\beta,\gamma,\theta,a,b,c,\phi) = \expn^{\iu \delta} \expn^{\iu  \alpha \lambda_3} \expn^{\iu \beta \lambda_2}\expn^{\iu \gamma \lambda_3} \expn^{\iu  \theta \lambda_5} \expn^{\iu  a\lambda_3} \expn^{\iu  b\lambda_2}  \expn^{\iu  c\lambda_3} \expn^{\iu  \phi\lambda_8}\,.
\end{equation}
We can see that $\det D = \expn^{3\iu\delta}$, which gives an extra $\mathrm{U(1)}$ phase as expected in the $\mathrm{U(3)}$ definition. Now, we would realize the $\mathrm{SU(3)}$ Hamiltonian numerically by mapping $D$ to the $U(3)$ elements.

For example, we take the numerical values of gauge fields as $\{\theta_1, \theta_2, \theta_3, \theta_4,\theta_5, \theta_6,\theta_7, \theta_8\}=  \{0.6, -0.3, 1.2 , 0.8,\\ -1.6 , 0.2, 0.9, 0.4    \}$ and $\omega_7/J =0.7$ in Fig.~1 and Fig.~3 in the main text. Then we can numerically solve for the nine parameters as $\{\delta, \alpha,\beta,\gamma,\theta,a,b,c,\phi\}= \{ 0.882, -0.925 , 0.199, 1.941, 1.032, -1.202, 0.658, 1.195, 1.644 \}$ up to the precision of three decimal places. Numerical calculation also finds $\mu_1 = 2.607, \mu_2= -0.303,\mu_3= -2.304$ such that $\mu_1 + \mu_2 + \mu_3 = 0$.

For the initial state of $\psi_n = (a_{n\rmred}, a_{n\rmgreen}, a_{n\rmblue})^T$, we have
\begin{equation}
\psi_n^\prime (t+t_\mathrm{R}) = \expn^{-iH_0t_\mathrm{R}}\psi_n = \mathbf{T}\psi_n (t)\,,
\end{equation}
where $t$ is some arbitrary time and $t_{\mathrm{R}}$ is the round-trip time, and the transfer matrix is defined by $\mathbf{T} = \expn^{-iH_0 t_{\mathrm{R}}}$. But since
$H_0 = U\tilde{H}_0 U^\dagger$, where $\tilde{H}_0$ is the diagonalized matrix as shown above. Using the fact that $V=UDU^\dagger$ implies $\expn^V = U\expn^D U^\dagger$, it follows that
\begin{equation}
\mathbf{T} = \expn^{-iH_0 t_{\mathrm{R}}} = U\expn^{-i\tilde{H}_0 t_\mathrm{R}} U^\dagger = U\begin{pmatrix}
\expn^{-i2Jt_\mathrm{R} \cos(k + \mu_1)} & 0 & 0\\
0 & \expn^{-i2Jt_\mathrm{R} \cos(k + \mu_2)} & 0\\
0 & 0 & \expn^{-i2Jt_\mathrm{R} \cos(k + \mu_3)}
\end{pmatrix} U^\dagger = U \tilde{\mathbf{T}}U^\dagger \,,
\end{equation}
where $
U= \expn^{\iu\delta} \expn^{\iu \alpha \lambda_3} \expn^{\iu \beta \lambda_2}\expn^{\iu \gamma \lambda_3} \expn^{\iu \theta \lambda_5} \expn^{\iu a\lambda_3} \expn^{\iu b\lambda_2}  \expn^{\iu c\lambda_3} \expn^{\iu \phi\lambda_8}\,.
$ Therefore, we actually have
\begin{equation} 
\mathbf{T} =  \expn^{\iu\delta} \expn^{\iu \alpha \lambda_3} \expn^{\iu \beta \lambda_2}\expn^{\iu \gamma \lambda_3} \expn^{\iu \theta \lambda_5} \expn^{\iu a\lambda_3} \expn^{\iu b\lambda_2}  \expn^{\iu c\lambda_3} \expn^{\iu \phi\lambda_8}  \tilde{\mathbf{T}} \expn^{-\iu \phi\lambda_8}\expn^{-\iu c\lambda_3} \expn^{-\iu b\lambda_2} \expn^{-\iu a\lambda_3}\expn^{-\iu \theta \lambda_5}\expn^{-\iu \gamma \lambda_3}\expn^{-\iu \beta \lambda_2}\expn^{-\iu \alpha \lambda_3}\expn^{-\iu\delta}\,.
\end{equation}
Notice that since the second Gell-Mann matrix can be decomposed as $\lambda_2  =\sigma_y \oplus 0$, hence
\begin{equation}
\expn^{i\beta \lambda_2} = \expn^{\iu\beta \sigma_y} \oplus 1 = V \expn^{\iu\beta \sigma_z} V^\dagger \oplus 1 = (V\oplus 1)( \expn^{\iu\beta \sigma_z} \oplus 1)(V^\dagger \oplus 1) \,.
\end{equation}
for $V= \expn^{-\iu\frac{\pi}{4}\sigma_z} \expn^{-\iu\frac{\pi}{4}\sigma_y}\,.$
Explicitly
\begin{equation}
\expn^{\iu\beta \lambda_2} = \begin{pmatrix}
\expn^{-\iu\frac{\pi}{4}} & 0 & 0\\
0 & \expn^{\iu\frac{\pi}{4}} & 0\\
0 & 0 & 1\\
\end{pmatrix}
\begin{pmatrix}
\cos \frac{\pi}{4} & \sin\frac{\pi}{4} & 0\\
- \sin\frac{\pi}{4} &  \cos\frac{\pi}{4} & 0\\
0 & 0 & 1 \\
\end{pmatrix}
\begin{pmatrix}
\expn^{\iu\beta} & 0 & 0\\
0 & \expn^{-\iu\beta} & 0\\
0 & 0 & 1\\
\end{pmatrix}
\begin{pmatrix}
\cos \frac{\pi}{4} & -\sin\frac{\pi}{4} & 0\\
\sin\frac{\pi}{4} &  \cos\frac{\pi}{4} & 0\\
0 & 0 & 1 \\
\end{pmatrix}
\begin{pmatrix}
\expn^{\iu\frac{\pi}{4}} & 0 & 0\\
0 & \expn^{-\iu\frac{\pi}{4}} & 0\\
0 & 0 & 1\\
\end{pmatrix}\,.
\end{equation}
Next, notice that the fifth Gell-Mann matrix is
\begin{equation}
\lambda_5 = \begin{pmatrix}
0 & 0 & -\iu\\
0 & 0 & 0 \\
\iu& 0 & 0\\
\end{pmatrix} \,.
\end{equation}
This follows that
\begin{equation}
\expn^{\iu\theta\lambda_5} = \begin{pmatrix}
\expn^{-\iu\frac{\pi}{4}} & 0 & 0\\
0 & 1 & 0\\
0 & 0 & \expn^{\iu\frac{\pi}{4}}\\
\end{pmatrix}
\begin{pmatrix}
\cos \frac{\pi}{4} & 0 & \sin\frac{\pi}{4} \\
0 & 1 & 0 \\
- \sin\frac{\pi}{4} & 0 & \cos\frac{\pi}{4} \\
\end{pmatrix}
\begin{pmatrix}
\expn^{\iu\theta} & 0 & 0\\
0 & 1 & 0\\
0 & 0 & \expn^{-\iu\theta}\\
\end{pmatrix}
\begin{pmatrix}
\cos \frac{\pi}{4} & 0 & - \sin\frac{\pi}{4} \\
0 & 1 & 0 \\
\sin\frac{\pi}{4} & 0 & \cos\frac{\pi}{4} \\
\end{pmatrix}
\begin{pmatrix}
\expn^{i\frac{\pi}{4}} & 0 & 0\\
0 & 1 & 0\\
0 & 0 & \expn^{-i\frac{\pi}{4}}\\
\end{pmatrix}\,.
\end{equation}
Let's define
\begin{equation}
V^\prime = \begin{pmatrix}
\expn^{-\iu\frac{\pi}{4}} & 0 & 0\\
0 & 1 & 0\\
0 & 0 & \expn^{\iu\frac{\pi}{4}}\\
\end{pmatrix}
\begin{pmatrix}
\cos \frac{\pi}{4} & 0 & \sin\frac{\pi}{4} \\
0 & 1 & 0 \\
- \sin\frac{\pi}{4} & 0 & \cos\frac{\pi}{4} \\
\end{pmatrix}\,,
\end{equation}
which is a retarder and rotator that acts on the first and third modes (the red (r) and blue (b) modes). 

Therefore, in full, we can realize the optical elements as follows:
\begin{equation}
\boxed{
\begin{aligned}
U &= \expn^{\iu\delta} (\expn^{\iu \alpha\sigma_z}\oplus 1 ) (V \expn^{\iu\beta \sigma_z} V^\dagger \oplus 1)(\expn^{\iu \gamma \sigma_z}\oplus 1 ) [V^\prime (\expn^{\iu\theta}\oplus 1 \oplus \expn^{-\iu\theta}) V^{\prime\dagger}]\\
&\quad\quad\quad\quad\quad\quad \times  (\expn^{\iu a\sigma_z}\oplus 1 ) (V \expn^{\iu b\sigma_z} V^\dagger \oplus 1 ) (\expn^{\iu c\sigma_z}\oplus 1 ) (\expn^{\iu\frac{\phi}{\sqrt{3}} I} \oplus \expn^{-\iu\frac{2\phi}{\sqrt{3}}})
\end{aligned}}\,.
\end{equation}
To enable color-orbit coupling, we can introduce the system with a coupler $\mathbf{T}_{\mathrm{coc}}=\expn^{\iu\tilde{\phi} \lambda_7}$, so the full transfer matrix becomes
\begin{equation}
     \mathbf{T}_{\mathrm{tot}} = \expn^{-\iu H_{\mathrm{tot}}(k)t_\mathrm{R}} = \expn^{-\iu ( H_0(k) - \omega_7 \lambda_7) t_\mathrm{R}} \approx \expn^{-\iu  H_0(k)  t_\mathrm{R}} \expn^{\iu \omega_7 t_{\mathrm{R}} \lambda_7} =  U\expn^{-\iu\tilde{H}_0 t_\mathrm{R}} U^\dagger \expn^{\iu \tilde{\phi}_7 \lambda_7} \equiv \mathbf{T}\cdot \mathbf{T_{\mathrm{coc}}}\,,
\end{equation}
where the approximation of the BCH identity similar to~\eqref{eq:BCHapprox} is allowed in the small $t_{\mathrm{R}}$ limit. Therefore
\begin{equation}
\boxed{
\begin{aligned}
    \mathbf{T}_{\mathrm{tot}} = (\expn^{\iu\delta} \expn^{\iu \alpha \lambda_3} \expn^{\iu \beta \lambda_2}\expn^{\iu \gamma \lambda_3} \expn^{\iu \theta \lambda_5} \expn^{\iu a\lambda_3} \expn^{\iu b\lambda_2}  \expn^{\iu c\lambda_3} \expn^{\iu \phi\lambda_8}  \tilde{\mathbf{T}} \expn^{-\iu \phi\lambda_8}\expn^{-\iu c\lambda_3} \expn^{-\iu b\lambda_2} \expn^{-\iu a\lambda_3}\expn^{-\iu \theta \lambda_5}\expn^{-\iu \gamma \lambda_3}\expn^{-\iu\beta \lambda_2}\expn^{-\iu \alpha \lambda_3}\expn^{-i\delta})\expn^{\iu\tilde{\phi} \lambda_7}
\end{aligned}}
\end{equation}
which realizes the color-orbit coupling Hamiltonian as
\begin{equation}
\boxed{
H_{\mathrm{tot}}(k) = 2J\cos(k+\theta_1 \lambda_1 + \theta_2 \lambda_2 + \cdots + \theta_8 \lambda_8) - \omega_7 \lambda_7}\,.
\end{equation}

\section{Simplified scheme for a subset $\mathrm{SU(3)}$ gauge group} \label{ExpSU3}
\begin{figure}
    \centering
    \includegraphics[scale=0.25]{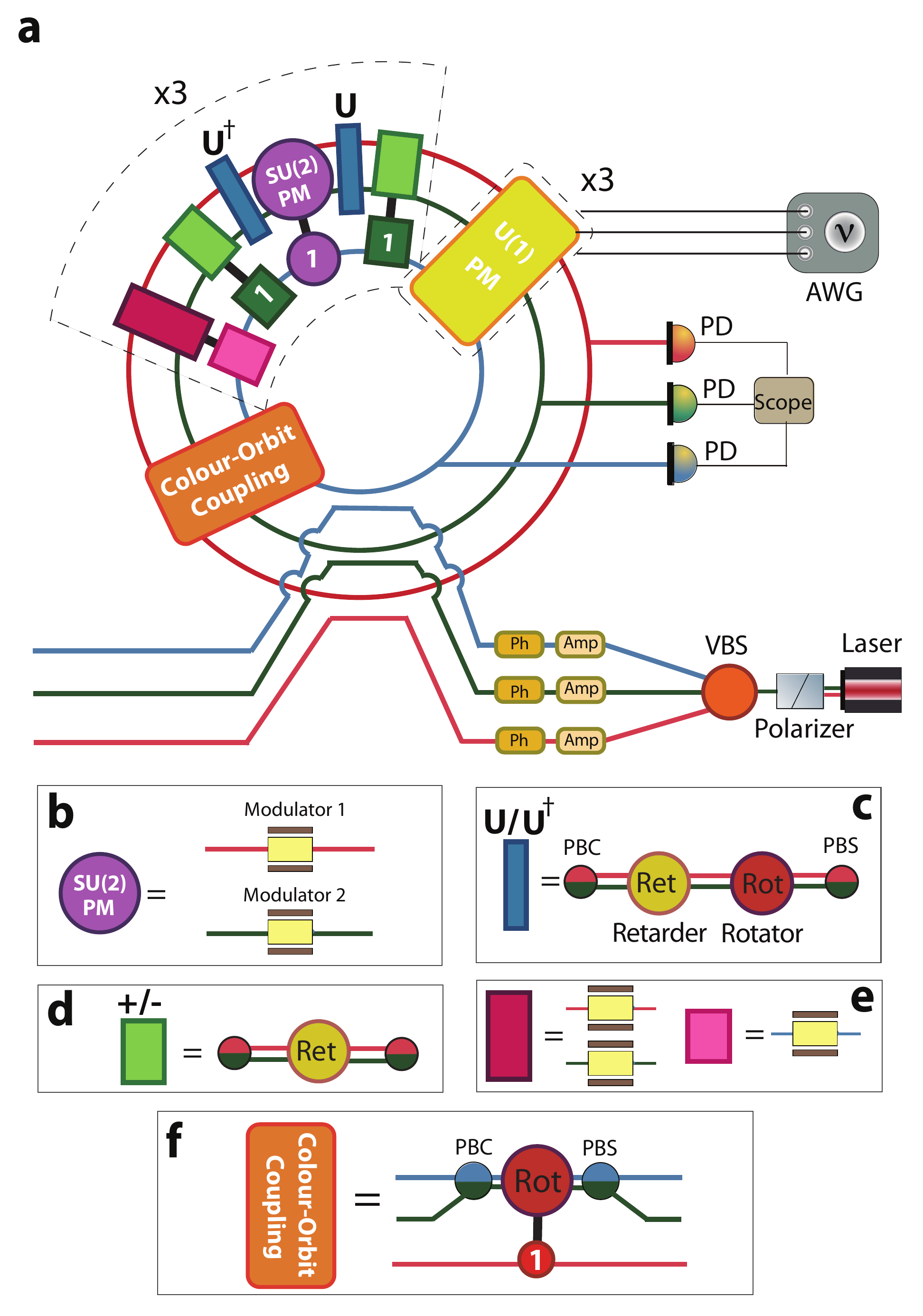}
    \centering
    \caption{\textbf{Photonic setup schematic to realize a subset of $\mathrm{SU(3)}$ gauge group.} \textbf{a.} Realization of a subset of $\mathrm{SU(3)}$ group with non-zero $\theta_1,\theta_2$ and $\theta_8$. Each colored component corresponds to each term in Eq.~\ref{eq:fullTransferFunction}. From right to left: Three identical $\mathrm{U(1)}$ phase modulations. Green boxes represent  $\expn^{\pm\iu \alpha_j \sigma_z} \oplus 1 $. Purple circle labeled with `SU(2) PM' and `1' sandwiched with two bluish green rectangles represent $\mathbf{U}\expn^{\iu 2J\tR\sqrt{\Theta_{j1}^2 + \Theta_{j2}^2}\cos(\Omega t + z_j \|\theta\|)\sigma_z}\mathbf{U}^\dagger \oplus 1$. The red and pink boxes represent $\expn^{-\iu \frac{2}{\sqrt{3}}J\tR \Theta_{j8} \cos(\Omega t + z_j \|\theta\|)\sigma_0} \oplus \expn^{\iu \frac{4}{\sqrt{3}}J\tR \Theta_{j8} \cos(\Omega t + z_j \|\theta\|)}$. 
    Finally, in addition we add the COC component $\expn^{\iu\tilde{\phi}\lambda_7} = 1\oplus \expn^{\iu\tilde{\phi}\sigma_y}$ to open the band gap. The two groups of optical elements are multiplied by three sets, due to three values of $z_1, z_2$ and $z_3$. 
    PBS/C: Polarization beam splitter/combiner; PD: Photodetector; VBS: Variable beam splitter. An input laser passing through a polarizer is connected to a variable beam splitter which splits into three identical parts. Then the three light pass through the amplitude modulator (Amp) followed by phase retarder (Ph) to form the desired arbitrary input states similar to the scheme in the main text, and enter the three rings in a clockwise manner.
    }
    \label{fig:setupfullpartial}
\end{figure}

Experimental realization of the full $\mathrm{SU(3)}$ could be challenging due to the complexity of the $\mathrm{SU(3)}$ group. Nonetheless, certain subsets of $\mathrm{SU(3)}$ can be more experimentally friendly. Here we develop the photonic set-up schematic for a partial $\mathrm{SU(3)}$ gauge group with an analytical closed form. From the three equations of motion in~\eqref{eq:EOM1}, ~\eqref{eq:EOM2} and ~\eqref{eq:EOM3}, or from the Hamiltonian~\eqref{eq:HamiltonianSU3}, we have
\begin{equation}
\iu \partial_t a_{n\alpha} = J\bigg(\expn^{-\iu(\theta_1\lambda_1 +\cdots  +\theta_1\lambda_8)}\bigg)_{\alpha\beta}a_{n+1\,\beta} + J\bigg(\expn^{\iu (\theta_1\lambda_1 +\cdots  +\theta_1\lambda_8)}\bigg)_{\alpha\beta}a_{n-1\,\beta}\,.  
\end{equation}
By Ref.~\cite{pang2024synthetic}, in matrix form, we have
\begin{equation}
a_{n}(T+1)-a_n(T) = -\iu \tR \bigg[J\expn^{-\iu (\theta_1\lambda_1 +\cdots  +\theta_1\lambda_8)} a_{n+1} +J \expn^{\iu (\theta_1\lambda_1 +\cdots  +\theta_1\lambda_8)} a_{n-1} \bigg]\,.
\end{equation}
Then it follows that 
\begin{equation*}
J\expn^{-\iu (\theta_1\lambda_1 +\cdots  +\theta_1\lambda_8)} a_{n+1} =  J\expn^{-\iu (\theta_1\lambda_1 +\cdots  +\theta_1\lambda_8)} \expn^{-\iu\Omega t}a_{n} \quad \text{and} \quad J\expn^{\iu(\theta_1\lambda_1 +\cdots  +\theta_1\lambda_8)} a_{n+1} =  J\expn^{\iu(\theta_1\lambda_1 +\cdots  +\theta_1\lambda_8)} \expn^{\iu\Omega t}a_{n}\,.
\end{equation*}
Therefore,
\begin{equation}
\begin{aligned}
&\quad a_{n}(T+1)-a_n(T) = -\iu\tR \bigg[J\expn^{-\iu(\Omega t + \theta_1\lambda_1 +\cdots  +\theta_1\lambda_8)} +J \expn^{\iu(\Omega t+\theta_1\lambda_1 +\cdots  +\theta_1\lambda_8)}   \bigg]a_n \\
&= -\iu \tR (2J)\bigg[\sum_{j=1}^3 P_j \cos(\Omega t + z_j \|\theta\|) + \sum_{j=1}^3 \cos(\Omega t + z_j \|\theta\|)\bigg(Q_j \frac{\theta_c}{\|\theta\|} +R_j d^{abc}\frac{\theta_a \theta_b}{\|\theta\|^2} \bigg)\lambda_c \bigg] a_n\,.
\end{aligned}
\end{equation}
Then we can model the scattering matrix  as,
\begin{equation}
\mathbf{T}= \left[\prod_{j=1}^3  \expn^{-\iu \tR \Gamma_j \cos(\Omega t + z_j \|\theta\|)} \right] \left[ \prod_{j=1}^3 \expn^{-\iu \tR \xi_j \cos(\Omega t + z_j \|\theta\|)}\right]\,,
\end{equation}
where
\begin{equation}
\Gamma_j = 2J P_j \quad \text{and} \quad \xi_j = 2J\sum_{c}\bigg( Q_j \frac{\theta_c}{\|\theta\|} + R_j \sum_{ab} d^{abc} \frac{\theta_a\theta_b}{\|\theta\|^2} \bigg)\lambda_c \equiv 2J \Theta_j\,,
\end{equation}
which is equivalent to Eq.\eqref{eq:transferSU3}. The first product consists of 3 copies of $\mathrm{U(1)}$ while the second product consists of pure SU(3) terms. Diagonalization of the SU(3) group element is unachievable, but we find that it can be factorized into products of factors. From Ref.~\cite{byrd1998differential}, one finds the SU(3) elements can be parametrized by Euler angles:
\begin{equation}
D(\alpha,\beta,\gamma,\theta,a,b,c,\phi) = \expn^{\iu \alpha \lambda_3}\expn^{\iu \beta \lambda_2}\expn^{\iu \gamma \lambda_3}\expn^{\iu \theta \lambda_5}\expn^{\iu a\lambda_3}\expn^{\iu b\lambda_2} \expn^{\iu c\lambda_3}\expn^{\iu \phi\lambda_8}\,.
\end{equation}
This gives the nine elements of the SU(3) matrices as
\begin{equation} \label{eq:209}
\begin{cases}
D_{11} &= \expn^{\iu \big(c+\frac{\phi}{\sqrt{3}} \big)}
\bigg[\expn^{\iu (a+\alpha + \gamma)}\cos b \cos \beta \cos\theta - \expn^{-\iu (a-\alpha + \gamma)} \sin b\sin\beta\bigg] \\
D_{12} &= \expn^{-\iu \big(c-\frac{\phi}{\sqrt{3}} \big)}
\bigg[\expn^{\iu (a+\alpha + \gamma)}\sin b \cos \beta \cos\theta  + \expn^{-\iu (a-\alpha + \gamma)} \cos b\sin\beta\bigg]\\
D_{13}&=\expn^{\iu \big(\alpha + \gamma - \frac{2\phi}{\sqrt{3}}\big)} \cos\beta \sin\theta\\
D_{21} &= \expn^{\iu \big(c+\frac{\phi}{\sqrt{3}} \big)}
\bigg[-\expn^{-\iu (a+\alpha + \gamma)}\sin b \cos \beta - \expn^{\iu (a-\alpha + \gamma)} \cos b\cos\theta \sin\beta\bigg]\\
D_{22} &= \expn^{-\iu \big(c-\frac{\phi}{\sqrt{3}} \big)}
\bigg[\expn^{-\iu (a+\alpha + \gamma)}\cos b \cos \beta   - \expn^{\iu (a-\alpha + \gamma)} \sin b\sin\beta\cos\theta\bigg]\\
D_{23}&=-\expn^{-\iu \big(\alpha - \gamma + \frac{2\phi}{\sqrt{3}}\big)} \sin\beta \sin\theta\\
D_{31}&=-\expn^{\iu \big(a +c+ \frac{\phi}{\sqrt{3}}\big)} \cos b \sin\theta\\
D_{32}&=-\expn^{\iu \big(a - c + \frac{\phi}{\sqrt{3}}\big)} \sin b \sin\theta\\
D_{33}&=\expn^{ -\frac{2\iu  \phi}{\sqrt{3}}} \cos\theta\\
\end{cases}\,.
\end{equation}
Here in this section, we consider a partial set of elements of $\{\lambda_1, \lambda_2, \lambda_8 \}$. Consider the following:
\begin{equation} \label{eq:210}
\expn^{-\iu \eta(\Theta_1 \lambda_1 + \Theta_2 \lambda_2 + \Theta_8\lambda_8)} = \begin{pmatrix}
\expn^{-\iu \frac{\Theta_8 \eta}{\sqrt{3}}}\cos(\sqrt{\Theta_1^2 + \Theta_2^2}\eta) & \frac{(-\iu \Theta_1 -\Theta_2)\expn^{-\iu \frac{\Theta_8 \eta}{\sqrt{3}}}\sin(\sqrt{\Theta_1^2 + \Theta_2^2}\eta)}{\sqrt{\Theta_1^2 + \Theta_2^2}} &0 \\
\frac{(-\iu \Theta_1 +\Theta_2)\expn^{-\iu \frac{\Theta_8 \eta}{\sqrt{3}}}\sin(\sqrt{\Theta_1^2 + \Theta_2^2}\eta)}{\sqrt{\Theta_1^2 + \Theta_2^2}}  & \expn^{-\iu \frac{\Theta_8 \eta}{\sqrt{3}}}\cos(\sqrt{\Theta_1^2 + \Theta_2^2}\eta)&0 \\
0 & 0 & \expn^{\frac{2\iu \Theta_8 \eta}{\sqrt{3}}}
\end{pmatrix}\,.
\end{equation}
Now we compare the matrix elements of this matrix to those of \ref{eq:209}. By comparing the terms, we demand $\theta=0$ and $b=0$, $c=0$. Next we demand $a+\alpha + \gamma =0$. This gives 
\begin{equation} \label{eq:211}
D(\alpha,\beta,\phi) = \begin{pmatrix}
\expn^{\frac{\iu\phi}{\sqrt{3}}}\cos\beta & \expn^{\frac{\iu \phi}{\sqrt{3}}} \expn^{2\iu\alpha} \sin\beta & 0 \\
-\expn^{\frac{\iu\phi}{\sqrt{3}}}\expn^{-2\iu\alpha}\sin\beta & \expn^{\frac{i\phi}{\sqrt{3}}} \cos\beta & 0 \\
0 & 0 & \expn^{-\frac{2\iu\phi}{\sqrt{3}}}
\end{pmatrix}\,.
\end{equation}
By comparing the two matrices in \ref{eq:211} and \ref{eq:210}, we get
\begin{equation}
\phi = -\Theta_8 \eta \quad\text{and}\quad \quad\beta= \sqrt{\Theta_1^2 + \Theta_2^2}\eta\,,
\end{equation}
where $\eta= 2J\tR\cos(\Omega t + z_j \|\theta\|) $. Furthermore, we have
\begin{equation}
\expn^{2\iu\alpha}\sin\beta = \cos 2\alpha \sin\beta + \iu \sin 2\alpha \sin\beta =\frac{-\iu\Theta_1 -\Theta_2}{\sqrt{\Theta_1^2 + \Theta_2^2}} \sin\Big( \sqrt{\Theta_1^2 + \Theta_2^2}\eta\Big)\,.
\end{equation}
By matching coefficients, this yields
\begin{equation}
\cos 2 \alpha = \frac{-\Theta_2}{\sqrt{\Theta_1^2 + \Theta_2^2}} \quad \text{and} \quad \sin 2 \alpha = \frac{-\Theta_1}{\sqrt{\Theta_1^2 + \Theta_2^2}} \,.
\end{equation}
This gives
\begin{equation}
\alpha = \frac{1}{2}\left(\tan^{-1}\bigg(\frac{\Theta_1}{\Theta_2}\bigg) + n\pi \right)\,.
\end{equation}
Hence, for $\theta_1,\theta_2,\theta_8 \neq 0$ while others are zero, one can check that the corresponding non-zero $\Theta$s are $\Theta_{1},\Theta_2, \Theta_8$ for a particular $j$. Thus
\begin{equation}
\begingroup\makeatletter\def\f@size{7.8}\check@mathfonts
\begin{aligned} 
& \expn^{-\iu 2J\tR\cos(\Omega t + z_j \|\theta\|)(\Theta_{j1} \lambda_1 + \Theta_{j2} \lambda_2 + \Theta_{j8}\lambda_8)} \\
&=\begin{pmatrix}
\expn^{-\iu \frac{2J\Theta_{j8} \tR\cos(\Omega t + z_j \|\theta\|)}{\sqrt{3}}}\cos(2J\tR\sqrt{\Theta_{j1}^2 + \Theta_{j2}^2}\cos(\Omega t + z_j \|\theta\|)) & \frac{(-\iu \Theta_{j1} -\Theta_{j2})\expn^{\frac{-\iu 2J\tR\Theta_{j8} \cos(\Omega t + z_j \|\theta\|)}{\sqrt{3}}}\sin(2J\tR\sqrt{\Theta_{j1}^2 + \Theta_{j2}^2}\cos(\Omega t + z_j \|\theta\|))}{\sqrt{\Theta_{j1}^2 + \Theta_{j2}^2}} &0 \\
\frac{(-\iu \Theta_{j1} +\Theta_{j2})\expn^{-i\frac{2J\tR\Theta_{j8} \cos(\Omega t + z_j \|\theta\|)}{\sqrt{3}}}\sin(2J\tR\sqrt{\Theta_{j1}^2 + \Theta_{j2}^2}\cos(\Omega t + z_j \|\theta\|))}{\sqrt{\Theta_{j1}^2 + \Theta_{j2}^2}}  & \expn^{-\iu \frac{2J\tR\Theta_{j8} \cos(\Omega t + z_j \|\theta\|)}{\sqrt{3}}}\cos(2J\tR\sqrt{\Theta_{j1}^2 + \Theta_{j2}^2}\cos(\Omega t + z_j \|\theta\|))&0 \\
0 & 0 & \expn^{\frac{4\iu J\tR\Theta_8 \cos(\Omega t + z_j \|\theta\|)}{\sqrt{3}}}
\end{pmatrix}\,.
\end{aligned}
\endgroup
\end{equation}
Finally, we have
\begin{equation}
D(\alpha,\beta,\phi)  =\expn^{\iu \alpha \lambda_3}\expn^{\iu \beta \lambda_2}\expn^{-\iu \alpha \lambda_3 } \expn^{\iu \phi\lambda_8}\,.
\end{equation}
Therefore, we see that the modulation by $\{\lambda_1, \lambda_2, \lambda_8 \}$ can be factorized into a product of 4 exponential Gell-Mann matrices. Thus
\begin{equation}
\expn^{-\iu 2J\tR\cos(\Omega t + z_j \|\theta\|)(\Theta_{j1} \lambda_1 + \Theta_{j2} \lambda_2 + \Theta_{j8}\lambda_8)} = \expn^{\iu \alpha_j \lambda_3}\expn^{\iu 2J \tR \sqrt{\Theta_{j1}^2 + \Theta_{j2}^2}\cos(\Omega t + z_j \|\theta\|)\lambda_2}\expn^{-\iu \alpha_j \lambda_3 } \expn^{-\iu 2J\tR \Theta_{j8} \cos(\Omega t + z_j \|\theta\|)\lambda_8}\,.
\end{equation}
The transfer matrix, in full, reads,
\begin{equation}
\mathbf{T} = \left[\prod_{j=1}^3 \expn^{-\iu 2J \tR P_j \cos(\Omega t + z_j \|\theta\|)}\right]\left[ \prod_{j=1}^3 \expn^{-\iu 2J\tR\cos(\Omega t + z_j \|\theta\|)(\Theta_{j1} \lambda_1 + \Theta_{j2} \lambda_2 + \Theta_{j8}\lambda_8)}\right]\,.
\end{equation}
Equivalently,
\begin{equation} \label{eq:213}
\mathbf{T} = \left[ \prod_{j=1}^3 \expn^{-\iu 2J \tR P_j \cos(\Omega t + z_j \|\theta\|)}\right]\left[\prod_{j=1}^3 \expn^{\iu \alpha_j \lambda_3}\expn^{\iu 2J\tR\sqrt{\Theta_{j1}^2 + \Theta_{j2}^2}\cos(\Omega t + z_j \|\theta\|)\lambda_2}\expn^{-\iu \alpha_j \lambda_3 } \expn^{-\iu 2J\tR \Theta_{j8} \cos(\Omega t + z_j \|\theta\|)\lambda_8}\right]\,.
\end{equation}
In our case, we only keep $\theta_1,\theta_2,\theta_8$ as non-zero. Then from \ref{eq:166}
\begin{equation}
\begin{cases}
\Theta_{j1} &= Q_j \frac{\theta_1}{\|\theta\|} +R_j d^{ab1}\frac{\theta_a\theta_b}{\|\theta\|^2} = Q_j \frac{\theta_1}{\|\theta\|} +R_j \frac{2}{\sqrt{3}}\frac{\theta_1\theta_8}{\|\theta\|^2} \\
\Theta_{j2} &= Q_j \frac{\theta_2}{\|\theta\|} +R_j d^{ab2}\frac{\theta_a\theta_b}{\|\theta\|^2} = Q_j \frac{\theta_2}{\|\theta\|} +R_j \frac{2}{\sqrt{3}}\frac{\theta_2\theta_8}{\|\theta\|^2} \\
\Theta_{j3} &=0\\
\Theta_{j4} &=0\\
\Theta_{j5} &=0\\
\Theta_{j6} &=0\\
\Theta_{j7} &=0\\
\Theta_{j8} &=Q_j \frac{\theta_8}{\|\theta\|} +R_j d^{ab8}\frac{\theta_a\theta_b}{\|\theta\|^2} = Q_j \frac{\theta_8}{\|\theta\|} +R_j \frac{1}{\sqrt{3}}\frac{\theta_1^2 + \theta_2^2 -\theta_8^2}{\|\theta\|^2} \\
\end{cases}
\end{equation}
and
\begin{equation}
\boxed{
\alpha_j = \frac{1}{2}\tan^{-1}\left(\frac{\Theta_{j1}}{\Theta_{j2}}\right) + \frac{n_j\pi}{2}=  \frac{1}{2}\tan^{-1}\left(\frac{Q_j \frac{\theta_1}{\|\theta\|} +R_j \frac{2}{\sqrt{3}}\frac{\theta_1\theta_8}{\|\theta\|^2} }{Q_j \frac{\theta_2}{\|\theta\|} +R_j \frac{2}{\sqrt{3}}\frac{\theta_2\theta_8}{\|\theta\|^2}}\right)+ \frac{n_j\pi}{2} = \frac{1}{2}\tan^{-1}\left(\frac{\theta_1}{\theta_2} \right)+ \frac{n_j\pi}{2}}\,,
\end{equation}
where the choice of $n_j \in \mathbb{Z}$ is subjected to the respective signs of $\Theta_{j1}$ and $\Theta_{j2}$ in order to match with the correct trigonometry quadrants. Let us investigate term by term in Eq.~\ref{eq:213} in terms of optics. Notice that the gamma matrix set $\{\lambda_1, \lambda_2, \lambda_8 \}$ we choose  can be written in terms of a direct sum:
\begin{equation}
\lambda_3 = \sigma_z \oplus 0 \quad\text{and} \quad \lambda_2 = \sigma_y \oplus 0 \quad\text{and} \quad \lambda_8 = \frac{1}{\sqrt{3}} [\sigma_0 \oplus (-2)] \,.
\end{equation} 
Therefore, we can write the individual exponential term as follows:
\begin{equation}
\expn^{\iu \alpha_j\lambda_3}=\expn^{\iu \alpha_j \sigma_z} \oplus 1 \,,
\end{equation}
\begin{equation}
\expn^{\iu 2J\tR\sqrt{\Theta_{j1}^2 + \Theta_{j2}^2}\cos(\Omega t + z_j \|\theta\|)\lambda_2} =\expn^{\iu 2J\tR\sqrt{\Theta_{j1}^2 + \Theta_{j2}^2}\cos(\Omega t + z_j \|\theta\|)\sigma_y} \oplus 1
\end{equation}
\begin{equation}
 \expn^{-\iu 2J\tR \Theta_{j8} \cos(\Omega t + z_j \|\theta\|)\lambda_8} =  \expn^{-\iu \frac{2}{\sqrt{3}}J\tR \Theta_{j8} \cos(\Omega t + z_j \|\theta\|)\sigma_0} \oplus \expn^{\iu \frac{4}{\sqrt{3}}J\tR \Theta_{j8} \cos(\Omega t + z_j \|\theta\|)} \,.
\end{equation}
Hence, we have the following schematic:
\begin{equation}
\begin{cases}
& \expn^{\iu \alpha_j \lambda_3}:\text{Retarder acting on the first two modes}\\
&\expn^{\iu 2J\tR\sqrt{\Theta_{j1}^2 + \Theta_{j2}^2}\cos(\Omega t + z_j \|\theta\|)\lambda_2}: \text{Rotator acting on the first two modes}\\
&\expn^{-\iu 2J\tR \Theta_{j8} \cos(\Omega t + z_j \|\theta\|)\lambda_8}: \text{Colour modulation for the last two modes}\\
\end{cases}\,.
\end{equation}
Therefore, the scattering matrix can be decomposed as
\begin{equation} 
\begin{aligned}
\mathbf{T} &=  \left[\prod_{j=1}^3 \expn^{-\iu 2J \tR P_j \cos(\Omega t + z_j \|\theta\|)}\right]\left[\prod_{j=1}^3 \expn^{\iu \alpha_j \lambda_3}\expn^{\iu 2J\tR\sqrt{\Theta_{j1}^2 + \Theta_{j2}^2}\cos(\Omega t + z_j \|\theta\|)\lambda_2}\expn^{-\iu \alpha_j \lambda_3 } \expn^{-\iu 2J\tR \Theta_{j8} \cos(\Omega t + z_j \|\theta\|)\lambda_8} \right]\\
&=\Bigg[\prod_{j=1}^3 \expn^{-\iu 2J \tR P_j \cos(\Omega t + z_j \|\theta\|)}\Bigg]\Bigg[\prod_{j=1}^3 (\expn^{\iu \alpha_j \sigma_z} \oplus 1 ) (\expn^{\iu 2J\tR\sqrt{\Theta_{j1}^2 + \Theta_{j2}^2}\cos(\Omega t + z_j \|\theta\|)\sigma_y} \oplus 1)   (\expn^{-\iu \alpha_j \sigma_z} \oplus 1 )\\
&\quad\quad\quad\quad\quad\quad\quad\quad\quad\quad\quad\quad\quad\quad  \left(\expn^{-\iu \frac{2}{\sqrt{3}}J\tR \Theta_{j8} \cos(\Omega t + z_j \|\theta\|)\sigma_0} \oplus \expn^{i\frac{4}{\sqrt{3}}J\tR \Theta_{j8} \cos(\Omega t + z_j \|\theta\|)}\right) \Bigg] \\
&= \left[\prod_{j=1}^3 \expn^{-\iu 2J \tR P_j \cos(\Omega t + z_j \|\theta\|)}\right]\\
&\quad\quad\times\left[\prod_{j=1}^3 \Big( \expn^{\iu \alpha_j \sigma_z}\expn^{\iu 2J\tR\sqrt{\Theta_{j1}^2 + \Theta_{j2}^2}\cos(\Omega t + z_j \|\theta\|)\sigma_y} \expn^{-\iu \alpha_j \sigma_z} \expn^{-\iu \frac{2}{\sqrt{3}}J\tR \Theta_{j8} \cos(\Omega t + z_j \|\theta\|)\sigma_0} \oplus \expn^{\iu \frac{4}{\sqrt{3}}J\tR \Theta_{j8} \cos(\Omega t + z_j \|\theta\|)}\Big)\right] \,.
\end{aligned}
\end{equation}
Furthermore, we need to transform the following term:
\begin{equation}
\expn^{\iu 2J\tR\sqrt{\Theta_{j1}^2 + \Theta_{j2}^2}\cos(\Omega t + z_j \|\theta\|)\sigma_y} = \mathbf{U}\expn^{\iu 2\tR \sqrt{\Theta_{j1}^2 + \Theta_{j2}^2}\cos(\Omega t + z_j \|\theta\|)\sigma_z}\mathbf{U}^\dagger\,,
\end{equation}
where we substitute
\begin{equation}
\theta_{xj} =0,\quad \theta_{zj}=0, \quad \theta_{yj} =2J \tR \sqrt{\Theta_{j1}^2 + \Theta_{j2}^2}\cos(\Omega t + z_j \|\theta\|) \,.
\end{equation}
It can be shown that the unitary matrix is given by
\begin{equation}
\mathbf{U} = R_z \bigg( \frac{\pi}{2} \bigg) R_y \bigg(  \frac{\pi}{2}\bigg) = \expn^{-\iu \frac{\pi}{4}\sigma_z } \expn^{-\iu \frac{\pi}{4}\sigma_y } \,.
\end{equation}
Therefore, the scattering matrix reads:
\begin{equation}
\begin{aligned}
\mathbf{T} &= \Bigg[\prod_{j=1}^3 \expn^{-\iu 2J \tR P_j \cos(\Omega t + z_j \|\theta\|)}\Bigg]\Bigg[\prod_{j=1}^3 (\expn^{\iu \alpha_j \sigma_z} \oplus 1 ) (\expn^{-\iu \frac{\pi}{4}\sigma_z } \expn^{- \iu \frac{\pi}{4}\sigma_y } \expn^{\iu 2J\tR\sqrt{\Theta_{j1}^2 + \Theta_{j2}^2}\cos(\Omega t + z_j \|\theta\|)\sigma_z}\expn^{ \iu \frac{\pi}{4}\sigma_y }\expn^{\iu \frac{\pi}{4}\sigma_z } \oplus 1)   (\expn^{-\iu \alpha_j \sigma_z} \oplus 1 )\\
&\quad\quad\quad\quad\quad\quad\quad\quad\quad\quad\quad\quad\quad\quad \times \left(\expn^{-\iu \frac{2}{\sqrt{3}}J\tR \Theta_{j8} \cos(\Omega t + z_j \|\theta\|)\sigma_0} \oplus \expn^{\iu \frac{4}{\sqrt{3}}J\tR \Theta_{j8} \cos(\Omega t + z_j \|\theta\|)} \right)\Bigg]\\
\end{aligned}\label{smeq:fullTransferFunction}\,.
\end{equation}
Together with the effective coupler $\mathbf{T}_{\mathrm{coc}}=\expn^{i\tilde{\phi} \lambda_7} = 1\oplus \expn^{i\tilde{\phi}\sigma_y}$, the full transfer matrix reads
\begin{equation}\label{eq:fullTransferFunction}
\boxed{
\begin{aligned}
\mathbf{T} &= \Bigg[\prod_{j=1}^3 \expn^{-\iu 2J \tR P_j \cos(\Omega t + z_j \|\theta\|)}\Bigg]\Bigg[\prod_{j=1}^3 (\expn^{\iu \alpha_j \sigma_z} \oplus 1 ) (\expn^{-\iu \frac{\pi}{4}\sigma_z } \expn^{- \iu \frac{\pi}{4}\sigma_y } \expn^{i2J\tR\sqrt{\Theta_{j1}^2 + \Theta_{j2}^2}\cos(\Omega t + z_j \|\theta\|)\sigma_z}\expn^{ \iu \frac{\pi}{4}\sigma_y }\expn^{\iu \frac{\pi}{4}\sigma_z } \oplus 1)   (\expn^{-\iu \alpha_j \sigma_z} \oplus 1 )\\
&\quad\quad\quad\quad\quad\quad\quad\quad\quad\quad\quad\quad\quad\quad \times \left(\expn^{-\iu \frac{2}{\sqrt{3}}J\tR \Theta_{j8} \cos(\Omega t + z_j \|\theta\|)\sigma_0} \oplus \expn^{\iu \frac{4}{\sqrt{3}}J\tR \Theta_{j8} \cos(\Omega t + z_j \|\theta\|)} \right)\Bigg](1\oplus \expn^{\iu\tilde{\phi}\sigma_y})\\
\end{aligned}}\,,
\end{equation}
where each element enclosed by the product can be realized as an individual optical element in the ring resonator. The setup schematic is illustrated in Fig. \ref{fig:setupfullpartial}.  The product of exponentials can be realized with different optical components sequentially aligned inside the resonators (see Fig.~\ref{fig:setupfullpartial}a), which includes phase modulation, polarization rotation, phase retardation, and polarization splitting and combining. 
Inside the ring resonator, we first include a combined component to realize color-orbit coupling $\mathbf{T}_{\mathrm{coc}}=\expn^{\iu\tilde{\phi}\lambda_7}$  where a Gell-Mann matrix $\lambda_7 = 0 \oplus \sigma_y$ with the phase $\tilde{\phi}=\omega_7 t_{\mathrm{R}}$. 

Within the second product sign, the term $\left(\expn^{-\iu \frac{2}{\sqrt{3}}J\tR \Theta_{j8} \cos(\Omega t + z_j |\theta|)\sigma_0} \oplus \expn^{\iu \frac{4}{\sqrt{3}}J\tR \Theta_{j8} \cos(\Omega t + z_j \|\theta\|)}\right) $ represents the phase modulation featuring identical modulation drive and the inner ring (blue ring), featuring a different drive of a doubled modulation depth and $\pi$ phase shift, respectively (Fig.~\ref{fig:setupfullpartial}e).

Next, $\left(\expn^{\iu \alpha_j \sigma_z} \oplus 1 \right)$ and $\left(\expn^{-\iu \alpha_j \sigma_z} \oplus 1 \right)$ denote a pair of opposite-signed retarders operated on the outer ring (red ring) only, and the term $\oplus 1$ denotes that no operation is performed on the inner blue ring (Fig.~\ref{fig:setupfullpartial}d). The retarders sandwich a pair of composite retarder-rotators realizing the functionality of $\mathbf{U}=\expn^{-\iu \frac{\pi}{4}\sigma_z } \expn^{ -\iu \frac{\pi}{4}\sigma_y }$ and $\mathbf{U}^\dagger$ as SU(2) unitary gauge transformations (Fig.~\ref{fig:setupfullpartial}c), which further sandwich an SU(2) phase modulation in the $\sigma_z$ basis $\expn^{\iu 2J\tR\sqrt{\Theta_{j1}^2 + \Theta_{j2}^2}\cos(\Omega t + z_j \|\theta\|)\sigma_z} \oplus 1 $ on the outer red ring (Fig.~\ref{fig:setupfullpartial}b). 
Finally, the terms in the first product sign represent three identical $\mathrm{U(1)}$ phase modulations across all colors. All product signs operate on the three root values $z_1, z_2$, and $z_3$.

The projected intensities of the subset of $\mathrm{SU(3)}$ group using Eq. 4 in the main text are plotted as follows:
\begin{figure}[H]
    \centering
    \includegraphics[width=1\linewidth]{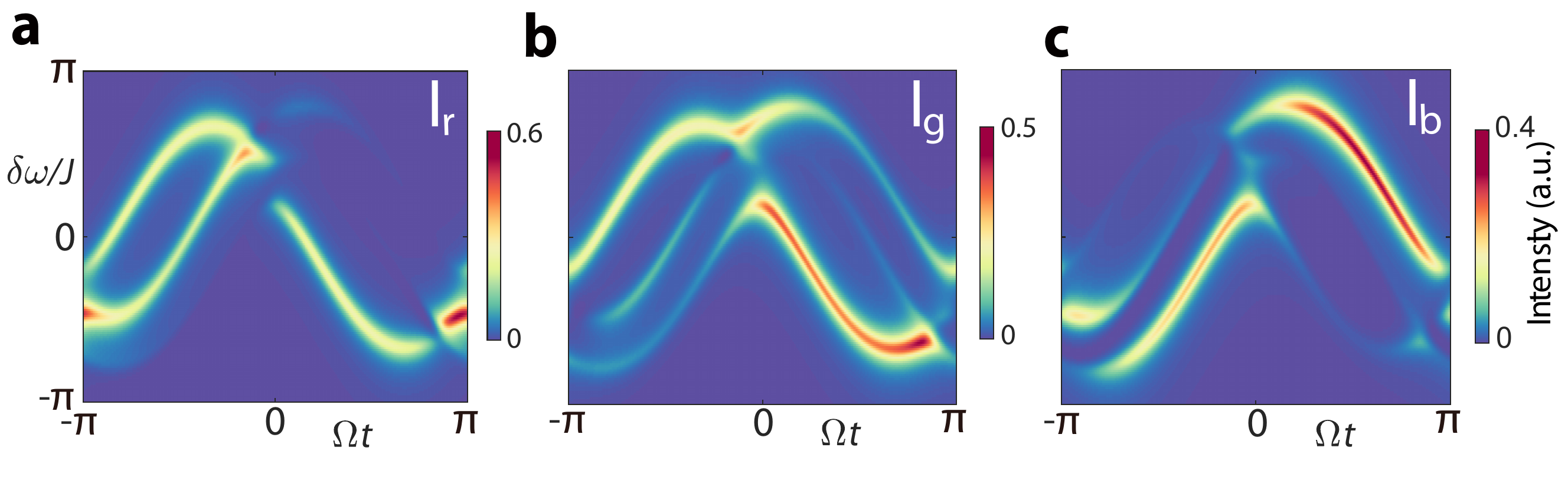}
    \centering
    \caption{\textbf{Projected intensities from the three rings for realizing a subset of the $\mathrm{SU(3)}$ gauge group.} From Fig.$\mathbf{a}$ to $ \mathbf{c}$, projected intensities of the first mode (red) $I_\mathrm{r}$, second mode $I_\mathrm{g}$ (green) and the third mode $I_\mathrm{b}$ (blue). The parameters used are same as above, $\{\theta_1 , \theta_2, \theta_8 \} =\{0.4, -0.5, 0.7 \}$\,rad., $\omega_7/J = 0.7$ and $\gamma_{\mathrm{in}}/J = 0.2$.
    } 
    \label{fig:SU3Fullintensity}
\end{figure}

\subsection{Transfer Matrix Simplification }
Here we provide an explicit simplification of the transfer matrix. Introducing back the $j$ index for the variables because we have three $z_j$\,s for $z_1 , z_2 $ and $z_3$, then the transfer matrix is explicitly given by
\begin{equation} 
\begin{aligned}
\mathbf{T} &= \left[ \prod_{j=1}^3 \expn^{-\iu 2J \tR P_j \cos(\Omega t + z_j \|\theta\|)}\right]\left[ \prod_{j=1}^3  D(\alpha_j,\beta_j,\phi_j) \right] \\
&=\left[ \prod_{j=1}^3 \expn^{-\iu 2J \tR P_j \cos(\Omega t + z_j \|\theta\|)} \right]\times\prod_{j=1}^3\begin{pmatrix}
\expn^{\frac{\iu\phi_j}{\sqrt{3}}}\cos\beta_j & \expn^{\frac{\iu \phi_j}{\sqrt{3}}} \expn^{2\iu\alpha_j} \sin\beta_j & 0 \\
-\expn^{\frac{\iu\phi_j}{\sqrt{3}}}\expn^{-2\iu\alpha_j}\sin\beta_j & \expn^{\frac{\iu\phi_j}{\sqrt{3}}} \cos\beta_j & 0 \\
0 & 0 & \expn^{-\frac{2\iu\phi_j}{\sqrt{3}}}
\end{pmatrix}\,.
\end{aligned}
\end{equation}
By direct calculation, the non-zero transfer matrix elements are given as
\begin{equation*}
\begin{aligned}
(T_H)_{11} &= P\expn^{\frac{\iu(\phi_1 + \phi_2 + \phi_3)}{\sqrt{3}}} \bigg( \cos\beta_1\cos\beta_2 \cos\beta_3 - \expn^{2\iu(\alpha_1 -\alpha_2)}\sin\beta_1 \sin\beta_2 \cos\beta_3 - \expn^{2\iu(\alpha_2 -\alpha_3)}\cos\beta_1 \sin\beta_2 \sin\beta_3  \\  
&\quad\quad\quad\quad\quad\quad\quad\quad\quad\quad\quad\quad\quad\quad\quad\quad - \expn^{2\iu(\alpha_1 -\alpha_3)}\sin\beta_1 \cos\beta_2 \sin\beta_3\bigg)
\end{aligned}
\end{equation*}
\begin{equation*}
\begin{aligned}
(T_H)_{12} &= P\expn^{\frac{\iu(\phi_1 + \phi_2 + \phi_3)}{\sqrt{3}}} \bigg( \expn^{2\iu\alpha_1}\sin\beta_1\cos\beta_2 \cos\beta_3 +\expn^{2\iu\alpha_2}\cos\beta_1 \sin\beta_2 \cos\beta_3 + \expn^{2\iu\alpha_3}\cos\beta_1 \cos\beta_2 \sin\beta_3  \\  
&\quad\quad\quad\quad\quad\quad\quad\quad\quad\quad\quad\quad\quad\quad\quad\quad - \expn^{2\iu(\alpha_1 -\alpha_2 + \alpha_3)}\sin\beta_1 \sin\beta_2 \sin\beta_3\bigg)
\end{aligned}
\end{equation*}
\begin{equation*}
\begin{aligned}
(T_H)_{21} &= P\expn^{\frac{\iu(\phi_1 + \phi_2 + \phi_3)}{\sqrt{3}}} \bigg( \expn^{-2\iu\alpha_1}\sin\beta_1\cos\beta_2 \cos\beta_3 +\expn^{-2\iu\alpha_2}\cos\beta_1 \sin\beta_2 \cos\beta_3 + \expn^{-2\iu\alpha_3}\cos\beta_1 \cos\beta_2 \sin\beta_3  \\  
&\quad\quad\quad\quad\quad\quad\quad\quad\quad\quad\quad\quad\quad\quad\quad\quad - \expn^{-2\iu(\alpha_1 -\alpha_2 + \alpha_3)}\sin\beta_1 \sin\beta_2 \sin\beta_3\bigg)
\end{aligned}
\end{equation*}
\begin{equation*}
\begin{aligned}
(T_H)_{22} &= P\expn^{\frac{\iu(\phi_1 + \phi_2 + \phi_3)}{\sqrt{3}}} \bigg( \cos\beta_1\cos\beta_2 \cos\beta_3 - \expn^{-2\iu(\alpha_1 -\alpha_2)}\sin\beta_1 \sin\beta_2 \cos\beta_3 - \expn^{-2\iu(\alpha_2 -\alpha_3)}\cos\beta_1 \sin\beta_2 \sin\beta_3  \\  
&\quad\quad\quad\quad\quad\quad\quad\quad\quad\quad\quad\quad\quad\quad\quad\quad - \expn^{-2\iu(\alpha_1 -\alpha_3)}\sin\beta_1 \cos\beta_2 \sin\beta_3\bigg)
\end{aligned}
\end{equation*}
\begin{equation}
\begin{aligned}
(T_H)_{33} &= P\expn^{\frac{-2\iu(\phi_1 + \phi_2 + \phi_3)}{\sqrt{3}}} \,,
\end{aligned}
\end{equation}
where $
  P= \prod_{j=1}^3 \expn^{-\iu 2J \tR P_j \cos(\Omega t + z_j \|\theta\|)} \,.$
In our parameter choices, for example, we pick $\theta_1 = 0.4$, $\theta_2 = -0.5$, $\theta_3 = 0.7$, we would have $\alpha_1 = \alpha_2$ and $\alpha_3 = \alpha_1 + \pi/2$. Furthermore, $P$ can be further simplified. Notice that
\begin{equation}
\begin{aligned}
P &=\prod_{j=1}^3\expn^{-\iu 2Jt_\mathrm{R} P_j (\cos\Omega t \cos(z_j\|\theta\|) -\sin\Omega t\sin(z_j\|\theta\|))}\\
&= \left[\prod_{j=1}^3      \expn^{-\iu 2Jt_\mathrm{R} P_j \cos\Omega t \cos(z_j\|\theta\|)  } \right] \left[ \prod_{j=1}^3 \expn^{\iu 2Jt_\mathrm{R} P_j \sin\Omega t \cos(z_j\|\theta\|)  } \right] \\
&= \expn^{-\iu 2J t_\mathrm{R} \big(\sum_{j=1}^3  \big(P_j\cos(z_j|  \theta|) \big) \cos\Omega t - \sum_{j=1}^3 \big(P_j\sin(z_j|  \theta|) \big) \sin\Omega t \big)     } \\
&= \expn^{-\iu 2Jt_\mathrm{R} R\cos(\Omega t + \gamma)} \,,
\end{aligned}
\end{equation}
where
\begin{equation}
    R= \sqrt{\Big( \sum_{j=1}^3 P_j\cos(z_j \|\theta\|)\Big)^2 + \Big( \sum_{j=1}^3 P_j \sin(z_j \|\theta\|)\Big)^2 } \quad \text{and} \quad \gamma  = \tan^{-1}\left( \frac{\sum_{j=1}^3 P_j\sin(z_j \|\theta\|)}{\sum_{j=1}^3 P_j\cos(z_j \|\theta\|)}   \right)\,.
\end{equation}
Then the transfer matrix simplifies to
\begin{equation}
\mathbf{T} =  \expn^{-\iu 2Jt_\mathrm{R} R\cos(\Omega t - \gamma)}\begin{pmatrix}
    \expn^{\frac{\iu (\phi_1 + \phi_2 + \phi_3)}{\sqrt{3}}} \cos(\beta_1 + \beta_2 - \beta_3) & \expn^{\frac{\iu (\phi_1 + \phi_2 + \phi_3)}{\sqrt{3}}} \expn^{2\iu\alpha_1} \sin(\beta_1 + \beta_2 -\beta_3) & 0 \\
\expn^{\frac{\iu (\phi_1 + \phi_2 + \phi_3)}{\sqrt{3}}} \expn^{-2\iu\alpha_1} \sin(\beta_1 + \beta_2 - \beta_3) & \expn^{\frac{\iu(\phi_1 + \phi_2 + \phi_3)}{\sqrt{3}}}\cos(\beta_1 + \beta_2 - \beta_3) & 0 \\
0 & 0 & \expn^{\frac{-2\iu(\phi_1 + \phi_2 + \phi_3)}{\sqrt{3}}} 
\end{pmatrix}\,.
\end{equation}

\section{Colour texture} \label{3DcolourTexture}

The color texture can be probed via transmission spectroscopy as in previous photonic synthetic frequency dimension works (e.g.~\cite{dutt2019experimental,yang2026non}).
Using the partial SU(3) group (SM. Sec. 4) as an example, we can calculate the color texture defined by the expectation values of $\lambda_1, \lambda_2$ and $\lambda_8$, which also depends on the effect of COC $\omega_7$:
\begin{equation}
    \langle \lambda_i (k ,\omega_7)\rangle_l = \frac{\langle \psi_l(k ,E_l (k) ,\omega_7) | \lambda_i | \psi_l(k ,E_l(k) ,\omega_7) \rangle }{\langle \psi_l(k ,E_l(k) ,\omega_7)| \psi_l(k ,E_l(k) ,\omega_7)\rangle}\,,
\end{equation}
where $i = 1,2,8$ and $l$ is the label of the energy band $1,2 \,\text{and}\,3$. The selected colour texture on Gell-Mann matrices $\lambda_1,\lambda_2$ and $\lambda_8$ for the subset of $\mathrm{SU(3)}$ group using the above gauge field parameters $\theta_1 = 0.4$, $\theta_2 = -0.5$, $\theta_8 = 0.7$ at $\omega_7/J = 0.7$ are presented in Fig. \ref{fig:SU3Fullcolourtexture} as follows:
\begin{figure}[H]
    \centering
    \includegraphics[width=1\linewidth]{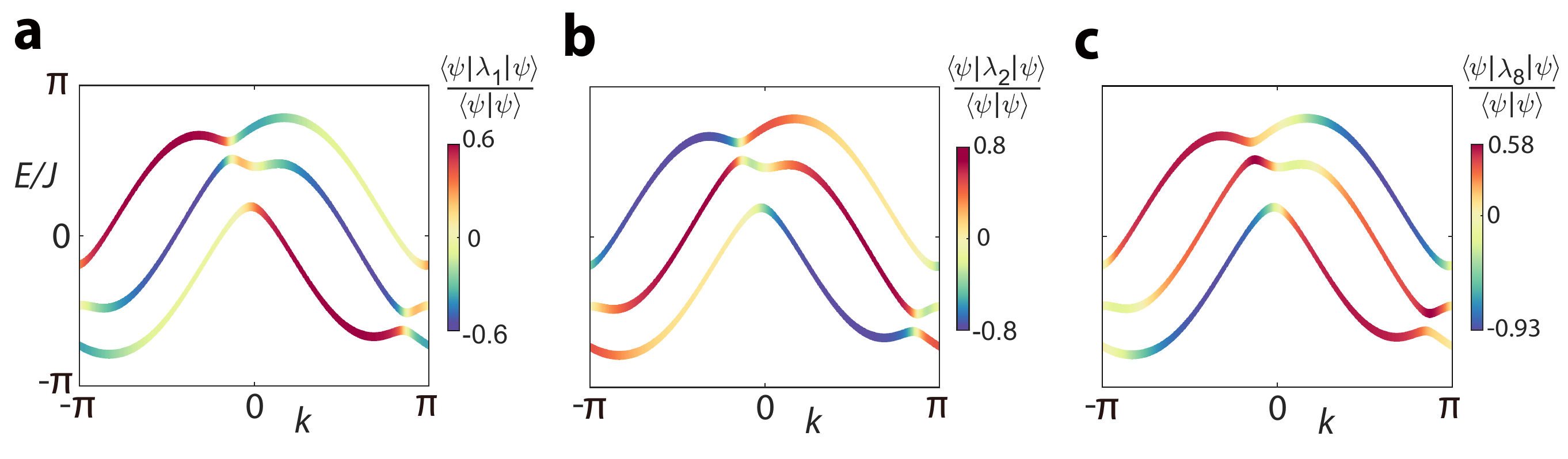}
    \centering
    \caption{
    \textbf{Colour texture of three eigen-energy bands for nonzero $\theta_1$, $\theta_2$ and $\theta_8$.}
    Colour texture of the subset of the $\mathrm{SU(3)}$ group with the chosen parameters above overlaid on the band structure for $\lambda_1$(Fig. $\mathbf{a}$), $\lambda_2$ (Fig. $\mathbf{b}$) and $\lambda_8$ (Fig. $\mathbf{c}$) are shown. The colour-orbit coupling strength is chosen as $\omega_7/J =0.7$. }
    \label{fig:SU3Fullcolourtexture}
\end{figure}
The colour texture with the same gauge field parameters above as a function of $\omega_7/J$ are presented as surface plots as follows in Fig. \ref{fig:SU3colourtexture3D}.
\begin{figure}[H]
    \centering
    \includegraphics[scale=0.22]{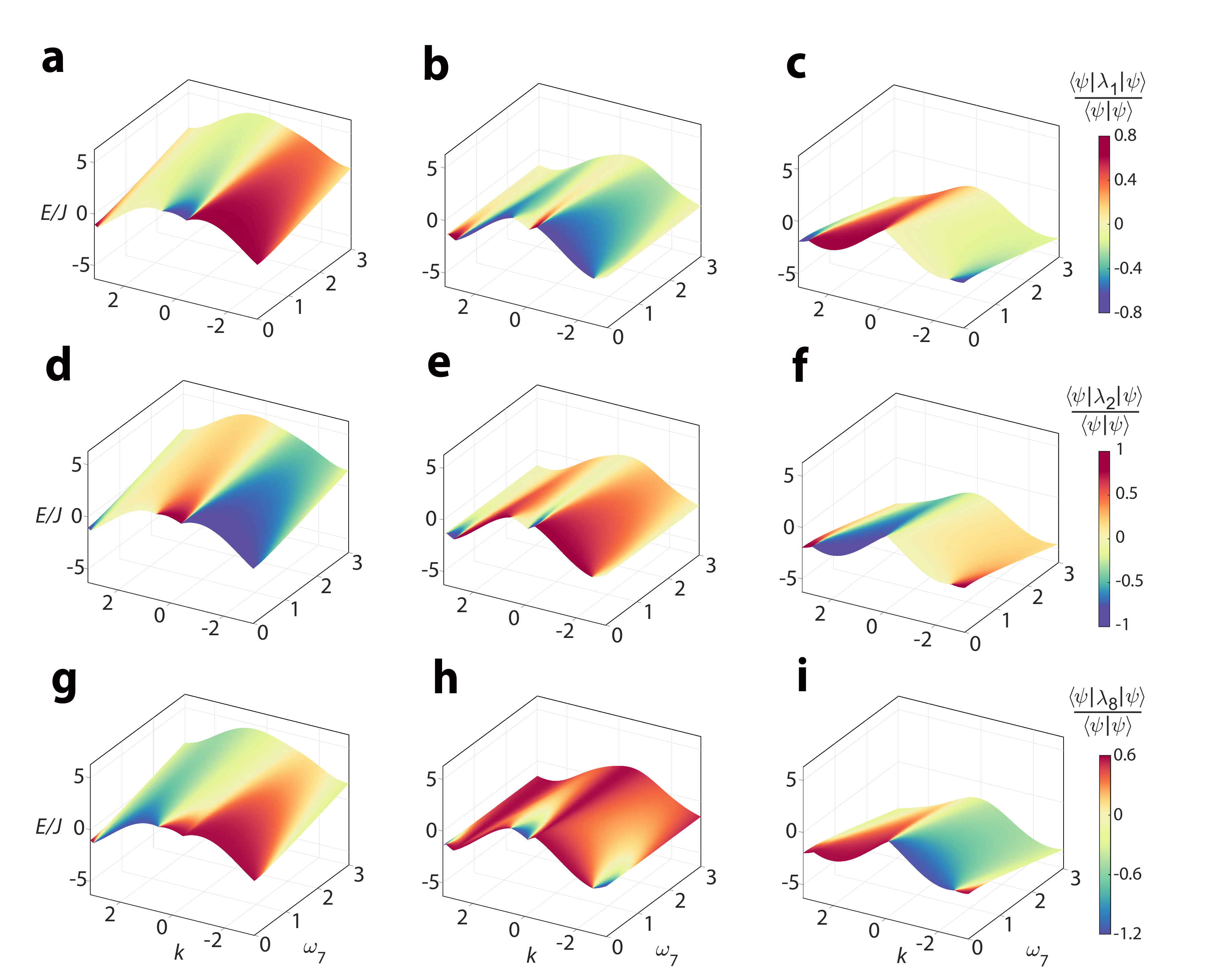}
    \centering
    \caption{
        \textbf{Surface plots of colour texture with different strength $\omega_7/J$ of colour-orbit coupling. }
    \textbf{a-c.} Colour texture  $\langle \lambda_1 (k ,\omega_7)\rangle_l$ with $l=3,2,1$ from left to right respectively. \textbf{d-f.} Colour texture  $\langle \lambda_2 (k ,\omega_7)\rangle_l$ with $l=3,2,1$ from left to right respectively. \textbf{g-i.} Colour texture  $\langle \lambda_8 (k ,\omega_7)\rangle_l$ with $l=3,2,1$ from left to right respectively. }
    \label{fig:SU3colourtexture3D}
\end{figure}

\section{Numerical Simulations}
We further validate our SU(3) scheme using numerical simulations based on Simulink\textsuperscript{®}. 
Driven by the numerical production of time-modulated transfer matrices, it is not limited by the first-order approximations as in the analytical derivations.
Moreover, the time-based simulation of the system enables it to handle the dynamic evolution of the signals, which are usually intractable to give an exact analytical solution. 

\begin{figure}[htbp]
    \centering
    \includegraphics[width=0.5\linewidth]{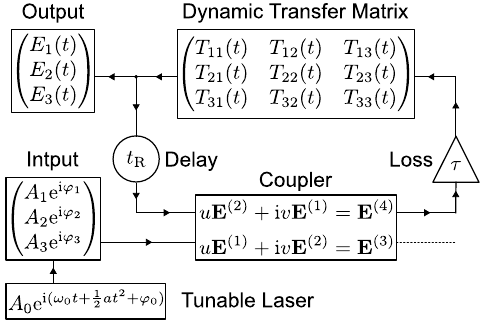}
    \caption{
    \textbf{Basic framework of the numerical simulation.}~The 3-dimensional signals flow along the arrowed solid lines. The boxes denote various operations on the signals.}
    \label{fig:SimulinkDiagram}
\end{figure}

A minimum version of the simulation setup is given in Fig.~\ref{fig:SimulinkDiagram}.
The signals propagating in the ring are three-dimensional vectors, which represent the field amplitudes in the three sub-rings in Fig.~\ref{fig:setupfullpartial} . 
We define the generalized coupler as a 4-port subsystem whose transfer function of the field amplitude is:
\begin{subequations} \label{smeq:coupler}
\begin{align}
u\pmb{E}^{(1)} + \mathrm{i} v\pmb{E}^{(2)} &= \pmb{E}^{(3)} \label{smeq:couplerP3}\\
u\pmb{E}^{(2)} + \mathrm{i} v\pmb{E}^{(1)} &= \pmb{E}^{(4)}  \,,\label{smeq:couplerP4}
\end{align}
\end{subequations}
where $u$ is the self-coupling coefficient, $v$ is the cross-coupling coefficient and $u^2+v^2=1$. $\pmb{E}^{(n)}$ $(n=1,2,3,4)$ is the field amplitude vector at the $n$-th port. 
The loss module and the time-delay module jointly model a lossy transport delay, and a programmable time-dependent transfer matrix summarizes all other modulations in the ring. 
Together, they achieve a round-trip transfer function of: 
\begin{equation}
\pmb{E}^{(2)}(t+t_\mathrm{R})=\tau\pmb{T}_H\times\pmb{E}^{(4)}(t) \,,
\end{equation}
where $\tau$ is the round-trip amplitude transmission factor, which is in general smaller than 1, and the full expression of the transfer matrix $\pmb{T}_H$ has been given in Eq.~\eqref{smeq:fullTransferFunction}.

The model in Fig.~\ref{fig:SimulinkDiagram} can be readily used to simulate the proposed system: we can always separate the signals into two or three sub-rings and decompose the complicated transfer matrix into minor elements such as rotators, retarders, polarization beam splitters, and phase modulators. 
In doing so, we recover the setup shown in Fig.~\ref{fig:setupfullpartial}, which is more favorable to experimental implementation, yet remains equivalent to the setup in Fig.~\ref{fig:SimulinkDiagram} at the simulation level.

To extract the band structure in the simulation, we applied the intensity projection method described in the main text by exciting the system with a tunable laser source, which is realized by a chirped harmonic oscillation with linearly increasing frequency: 
\begin{equation}
\pmb{E}_\mathrm{in}(t)=A_0\mathrm{e}^{\mathrm{i}(\omega_0t+\frac{1}{2}at^2+\varphi_0)}\,,
\end{equation}
where $A_0$ is the field amplitude, $\omega_0$ is the initial frequency, $a$ is the linear frequency-sweeping rate that leads to a frequency detuning of $\delta\omega=at$, and $\varphi_0$ is the initial phase. 
Here, we have chosen the convention of $+\mathrm{i}$ to match the theoretical analysis. 
The input state is prepared by multiplying the source by a state vector, such that:
\begin{equation}
\pmb{E}^{(1)}(t)=
\begin{pmatrix}
A_1\mathrm{e}^{\mathrm{i}\varphi_1} \\
A_2\mathrm{e}^{\mathrm{i}\varphi_2} \\
A_3\mathrm{e}^{\mathrm{i}\varphi_3}
\end{pmatrix}
\times\pmb{E}_\mathrm{in}(t) \,.
\end{equation}

\begin{figure}[htbp]
    \centering
    \includegraphics[width=1\linewidth]{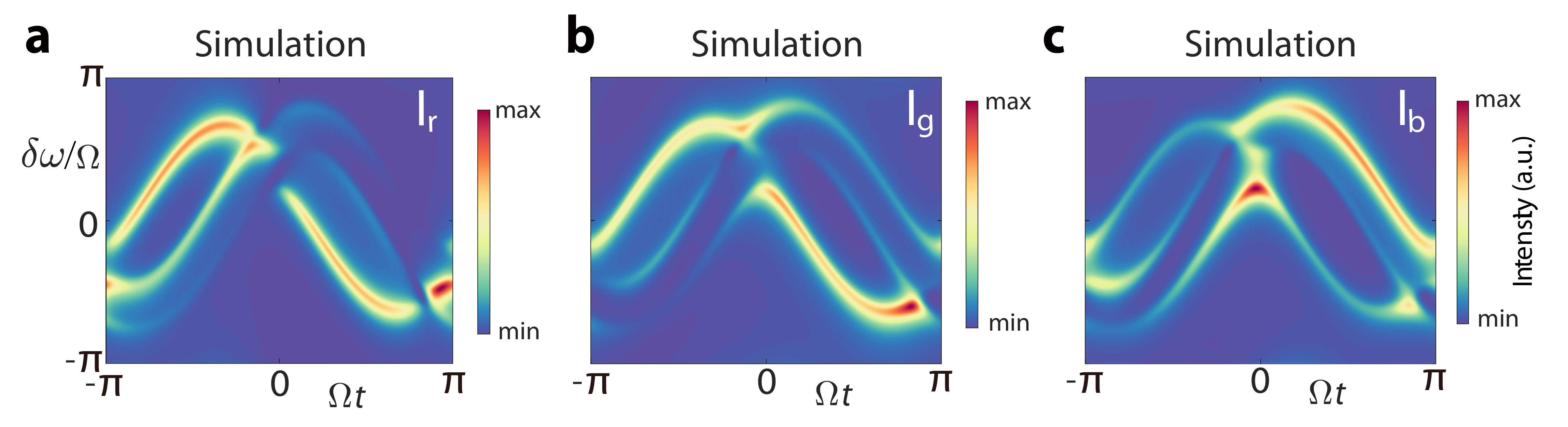}
    \centering
    \caption{\textbf{Numerically simulated intensity projection of photonic bands under synthetic SU(3) gauge fields.} Parameters are set to be consistent with the part of colour texture in Fig.~\ref{fig:SU3Fullcolourtexture}.}
    \label{fig:SU3simulink}
\end{figure}

Fig.~\ref{fig:SU3simulink} shows the numerical simulation of the supplementary material  Fig.~~\ref{fig:SU3Fullintensity}, which is in good agreement with the first-order analytical approximation. 
Most of the features of the band structure are retained, but the relative intensities in some parts of the intensity projection map slightly deviate from the theoretical prediction.
We attribute the minor discrepancies to the first-order and nearest-coupling approximations that limit the accuracy of analytical solutions.

\section{SU(3) Dynamics analysis}
\subsection{Case 1: without color-orbit coupling}
By using the Hamiltonian $H(k) = 2J\cos(k + \theta_1\lambda_1 +\theta_2 \lambda_2 +\theta_8\lambda_8) -\omega_7 \lambda_7$, for the simplest case, we first turn off the color-orbit coupling such that we set $\omega_7 =0$; we can compute the wave function quantum mechanically by the following equation:
\begin{equation}
\psi(x,t) = \frac{1}{\sqrt{2\pi}}\int_{-\infty}^\infty dk\, \expn^{-\iu H(k)t}\expn^{\iu kx} \langle k | \psi(0)\rangle \,.
\end{equation}
By direct computation, one finds
\begin{equation}
\expn^{-\iu H(k)t} =  
\begin{pmatrix}
   A(k,t)\cos\Big( 2Jt\sin\sqrt{\theta_1^2 +\theta_2^2 } \sin \big(k+\frac{\theta_8}{\sqrt{3}}\big)\Big) & A(k,t)\frac{\iu(\theta_1 + \iu\theta_2)}{\sqrt{\theta_1^2 +\theta_2^2 }}\sin\Big( 2Jt \sqrt{\theta_1^2 +\theta_2^2 }\sin \big(k+\frac{\theta_8}{\sqrt{3}} \big)\Big) & 0 \\
   A(k,t)\frac{\iu(\theta_1 - \iu\theta_2)}{\sqrt{\theta_1^2 +\theta_2^2 }}\sin\Big(2 Jt \sqrt{\theta_1^2 +\theta_2^2 }\sin \big( k+\frac{\theta_8}{\sqrt{3}} \big)\Big) & A(k,t)\cos\Big(2 Jt\sin\sqrt{\theta_1^2 +\theta_2^2 } \sin \big(k+\frac{\theta_8}{\sqrt{3}}\big)\Big) & 0\\
   0 & 0 & \expn^{-2iJt \cos\big(k - \frac{2\theta_8}{\sqrt{3}} \big)}
\end{pmatrix}\,,
\end{equation}
where $A(k,t)=\expn^{-2\iu Jt \sqrt{\theta_1^2 +\theta_2^2 }\cos\big(k+\frac{\theta_8}{\sqrt{3}}\big)}$.
Then we pick the initial Gaussian input state as 
\begin{equation}\label{eq:initialstate}
    \psi(k,0) = \langle k |\psi(0) \rangle = \frac{1}{\sqrt{2}}\bigg(\frac{2\gamma}{\pi}\bigg)^{1/4}\expn^{-\gamma k^2} \begin{pmatrix}
        1\\
        0\\
        1\\
    \end{pmatrix} \,.
\end{equation}
Then it directly follows that the three components of the wave function are:
\begin{equation} \label{eq:wavefunctionNoOmega}
    \begin{cases}
    \psi_{\mathrm{r}}(x,t) &= \frac{1}{2\sqrt{\pi}}\big(\frac{2\gamma}{\pi}\big)^{1/4}\int_{-\infty}^\infty \diff k\,\expn^{-2\iu Jt\sqrt{\theta_1^2 + \theta_2^2}\cos\big(k+ \frac{\theta_8}{\sqrt{3}}\big)+ \iu kx - \gamma k^2  }\cos\Big(2 Jt \sin\sqrt{\theta_1^2 + \theta_2^2} \sin\big(k+ \frac{\theta_8}{\sqrt{3}}\big) \Big) \\
     \psi_{\mathrm{g}}(x,t) &= \frac{1}{2\sqrt{\pi}}\big(\frac{2\gamma}{\pi}\big)^{1/4}\int_{-\infty}^\infty \diff k\,\expn^{-2\iu Jt\sqrt{\theta_1^2 + \theta_2^2}\cos\big(k+ \frac{\theta_8}{\sqrt{3}}\big)+ \iu kx - \gamma k^2  }\frac{\iu (\theta_1 -\iu \theta_2)}{\sqrt{\theta_1^2 + \theta_2^2}}\sin\Big(2 Jt \sin\sqrt{\theta_1^2 + \theta_2^2} \sin\big(k+ \frac{\theta_8}{\sqrt{3}}\big) \Big)\\
     \psi_{\mathrm{b}}(x,t) &= \frac{1}{2\sqrt{\pi}}\big(\frac{2\gamma}{\pi}\big)^{1/4}\int_{-\infty}^\infty \diff k\, \expn^{-2\iu Jt \cos\big( k-\frac{2\theta_8}{\sqrt{3}} \big) + \iu k x -\gamma k^2 }
    \end{cases}\,.
\end{equation}
The total intensity is then given by
\begin{equation}
I_{\mathrm{tot}} = |\psi_{\mathrm{r}}(x,t)|^2 + |\psi_{\mathrm{g}}(x,t)|^2 + |\psi_{\mathrm{b}}(x,t)|^2.
\end{equation}
The center of mass is then calculated by
\begin{equation}
    \langle x(t) \rangle = \int_{-\infty}^{\infty} xI_{\mathrm{tot}} \diff x\,.
\end{equation}

\subsection{Case 2: with color-orbit coupling}
For the general case, when we turn on the color-orbit coupling, i.e. $\omega_7 \neq 0$, the analysis requires a much more complicated and advanced technique. The computation of $\expn^{-iH(k)t}$ is not straightforward. In order to obtain an analytical expression, we need to again employ Sylvester's theorem as in Sec. 2 above. We have
\begin{equation}
    \expn^{-\iu H(k)t} = \sum_{j=1}^3 \expn^{-\iu E_j t} \prod_{m \neq j} \frac{H(k) - E_m(k) I}{E_j(k) - E_m(k)}\,.
\end{equation}
Now, let us rewrite our original partial $\mathrm{SU(3)}$ in a simpler form. Let $r =\sqrt{\theta_1^2 + \theta_2^2}$, $K(k) = k + \frac{\theta_8}{\sqrt{3}}$, $a(k)=\cos K \cos r$, $p(k) = \sin K \sin r$ and $b(k) = \cos\big( k - \frac{2\theta_8}{\sqrt{3}}\big)$. Then $H(k)$ now reads,
\begin{equation}
    H(k) =\begin{pmatrix}
       2Ja(k)  & -2J\frac{\theta_1 -\iu\theta_2}{r}p(k)  &0\\
       -2J\frac{\theta_1 +\iu\theta_2}{r}p(k)  &2Ja(k)  & \iu \omega_7 \\
       0 &-\iu \omega_7 & 2Jb(k) 
    \end{pmatrix} \,.
\end{equation}
We would like to further simplify the expression by defining a similarity transformation on $H(k)$. Define
\begin{equation}
    \mathbf{U } = \begin{pmatrix}
        \frac{\theta_1 + \iu\theta_2}{r} & 0 &0 \\
        0 & 1 & 0\\
        0 & 0 & \iu
    \end{pmatrix} 
\end{equation}
such that we can define the reduced Hamiltonian
\begin{equation}
    \tilde{H}(k) : = \mathbf{U}H(k) \mathbf{U}^\dagger = \begin{pmatrix}
        A(k)  & -B(k)  & 0 \\
        -B(k)  & A(k)  & \omega_7 \\
        0 & \omega_7 & C(k) 
    \end{pmatrix} \,,
\end{equation}
where $A=2Ja(k) $, $B=2Jp(k) $ and $C= 2Jb(k) $. Furthermore, let $\{\tilde{E}_0(k) , \tilde{E}_1(k) , \tilde{E}_2(k) \}$ be the three non-degenerate eigenvalues of $\tilde{H}(k)$. Therefore, the phase factor $ \expn^{-\iu H(k)t}$ can be obtained by
\begin{equation}
     \expn^{-\iu H(k)t} = \mathbf{U}^\dagger  \expn^{-\iu \tilde{H}(k)t} \mathbf{U} = \mathbf{U}^\dagger \left(\sum_{j=0,1,2} \expn^{-\iu\tilde{E}_j(k) t} \prod_{m \neq j} \frac{\tilde{H}(k) - \tilde{E}_m(k) I}{\tilde{E}_j(k) - \tilde{E}_m(k)}    \right) \mathbf{U}\,.
\end{equation}
The remaining task is to find the three eignevalues $\{\tilde{E}_0(k) , \tilde{E}_1(k) , \tilde{E}_2(k) \}$ by solving the characteristic polynomial. Consider $\det\big(\tilde{H}(k) - \tilde{E}(k) I \big) =0$, then the characteristic polynomial yields $ (C-\tilde{E})\big( (A-\tilde{E})^2 - B^2 \big) - (A-\tilde{E})\omega^2_7 =0 $. To simplify the expression, let $\delta = C- A$, $x= A-\tilde{E}$, then we obtain $x^3 + \delta x^2 -(B^2 + \omega_7^2)x - \delta B^2 = 0$. Furthermore, by substituting $x=y-\delta/3$, this gives us $y^3 + Py + Q =0$ with $P=-(B^2 +\omega_7^2) -\delta^2/3$ which is smaller than zero, and $Q = 2\delta^3/27 + \delta (\omega_7^2 -2B^2)/3$. The solution to the cubic equation is
\begin{equation}
    \tilde{E}_j = \frac{2A(k)  +C(k) }{3} - y_j(k)  \,,
\end{equation}
where 
\begin{equation}
    y_j(k)  =2\sqrt{\frac{-P(k) }{3}}\cos\left( \frac
    {\phi(k) -2\pi j}{3}\right) \quad \mathrm{with} \quad \phi(k)  = \cos^{-1}\left( \frac{3Q(k) }{2P(k) }\sqrt{\frac{-3}{P(k) }}\right)
\end{equation}
for $  j= 0,1,2$. Explicitly, we have the phase factor as
\begin{equation}
\begin{aligned}
     \expn^{-\iu H(k)t}  &= \mathbf{U}^\dagger \Bigg(\expn^{-\iu\tilde{E}_0 (k)t} \frac{(\tilde{H}(k)-\tilde{E}_1(k)I)(\tilde{H}(k)-\tilde{E}_2(k)I)}{(\tilde{E}_0(k)-\tilde{E}_1(k))(\tilde{E}_0(k)-\tilde{E}_2(k))} + \expn^{-\iu\tilde{E}_1 (k)t} \frac{(\tilde{H}(k)-\tilde{E}_0(k)I)(\tilde{H}(k)-\tilde{E}_2(k)I)}{(\tilde{E}_1(k)-\tilde{E}_0(k))(\tilde{E}_1(k)-\tilde{E}_2(k))} \\&\quad\quad\quad\quad\quad\quad\quad\quad\quad\quad\quad\quad\quad\quad\quad\quad + \expn^{-\iu\tilde{E}_2 (k)t} \frac{(\tilde{H}(k)-\tilde{E}_0(k)I)(\tilde{H}(k)-\tilde{E}_1(k)I)}{(\tilde{E}_2(k)-\tilde{E}_0(k))(\tilde{E}_2(k)-\tilde{E}_1(k))}
     \Bigg) \mathbf{U}\,.
     \end{aligned}
\end{equation}
Hence, compactly, the wave function is
\begin{equation} \label{eq:psisquareomega}
\boxed{
\psi(x,t) = \frac{1}{\sqrt{2\pi}}\int_{-\infty}^\infty \diff k\, \mathbf{U}^\dagger \left(\sum_{j=0,1,2} \expn^{-\iu \tilde{E}_j t} \prod_{m \neq j} \frac{\tilde{H}(k) - \tilde{E}_m(k) I}{\tilde{E}_j(k) - \tilde{E}_m(k)}    \right)\mathbf{U} \expn^{\iu kx} \langle k | \psi(0)\rangle} \,.
\end{equation}
The position center of mass is then
\begin{equation}
\boxed{
    \langle x(t)\rangle = \frac{1}{2\pi} \int_{-\infty}^\infty \diff x\,x \bigg\vert \int_{-\infty}^\infty \diff k\, \mathbf{U}^\dagger \left(\sum_{j=0,1,2} \expn^{-\iu \tilde{E}_j t} \prod_{m \neq j} \frac{\tilde{H}(k) - \tilde{E}_m(k) I}{\tilde{E}_j(k) - \tilde{E}_m(k)}    \right)\mathbf{U} \expn^{\iu kx} \langle k | \psi(0)\rangle \bigg\vert^2}\,.
\end{equation}
We will pick the initial state given in \ref{eq:initialstate} in all of our analysis. We have also numerically checked that the above method by the calculation of Eq.\eqref{eq:psisquareomega} is able to reduce to the method in Eq.\ref{eq:wavefunctionNoOmega} for the $\omega_7 = 0$ case. The plots for intensity and position expectation value for $\omega_7=0$ and small $\omega_7$ (here we choose $\omega_7= 0.2\,\mathrm{rad\,s^{-1}}$) are shown as follows.
\begin{figure}[H]
    \centering
    \includegraphics[width=1\linewidth]{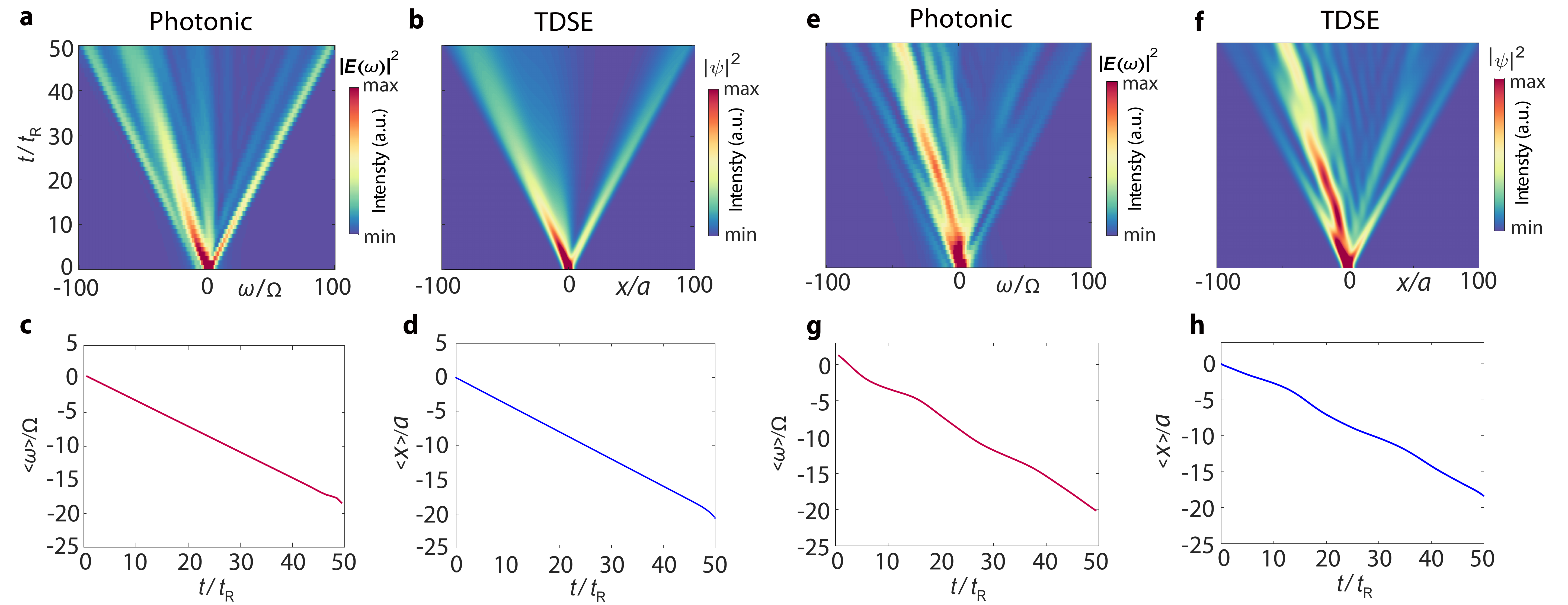}
    \centering
    \caption{\textbf{Comparison of SU(3) dynamics calculated using photonic simulations and the time-dependent Schrodinger equation (TDSE).} \textbf{a-d}. No colour-orbit coupling, $\omega_7 =0$. The center of mass moves towards the left in a unidirectional manner; \textbf{e-h}. Small wiggles along the unidirectional trajectory begin to appear by introducing a small $\omega_7 =0.2$, for which the dynamics of Zitterbewegung (ZB) start to take shape.}
    \label{fig:SU3dynamics}
\end{figure}
For both cases, the simulation and theoretical TDSE calculation match well. When $\omega_y$ becomes larger, the feature of Zitterbewegung (ZB) becomes more apparent (see Fig. 4 in the main text).

%\bibliographystyle{apsrev4-2}
%\bibliography{sources0}

%